\documentclass[aps,amsmath,amssymb,twocolumn]{revtex4-2}

\usepackage[]{graphicx}
\usepackage{color}
\newcommand{\be}{\begin{equation}}
\newcommand{\ee}{\end{equation}}
\newcommand{\ben}{\begin{eqnarray}}
\newcommand{\een}{\end{eqnarray}}
\newcommand{\bes}{\begin{subequations}}
\newcommand{\ees}{\end{subequations}}

\newcommand{\nn}{\nonumber\\}
\newcommand{\bfi}{\begin{figure}}
\newcommand{\efi}{\end{figure}}
\newcommand{\bc}{\begin{center}}
\newcommand{\ec}{\end{center}}
\newcommand{\sech}{\mbox{sech}}

\begin{document}
\title{Effects of Cuscuton dynamics on braneworld configurations \\in the scalar-tensor representation of $f\left(R,T\right)$ gravity }
\author{Jo\~ao Lu\'is Rosa$^{1}$, D. Bazeia$^{2}$ and A. S. Lob\~ao Jr.$^{3}$}
\affiliation{$^1$Institute of Physics, University of Tartu, W. Ostwaldi 1, 50411 Tartu, Estonia\\
$^2$Departamento de F\'\i sica, Universidade Federal da Para\'\i ba, 58051-970 Jo\~ao Pessoa, PB, Brazil\\
$^3$Escola T\'ecnica de Sa\'ude de Cajazeiras, Universidade Federal de Campina Grande, 58900-000 Cajazeiras, PB, Brazil\\
}
\begin{abstract}
In this work we study braneworld configurations in the scalar-tensor representation of the $f\left(R,T\right)$ gravity theory in the presence of a Cuscuton term in the source field matter Lagrangian. We start by deriving the scalar-tensor representation of the theory and obtaining its respective equations of motion. We then introduce the standard metric for a flat braneworld model with one extra dimension for the general case of a $f\left(R,T\right)$ theory and study two different models for the source field. Given the complexity of the field equations, these solutions are obtained numerically. In these cases, the Cuscuton term effectively amplifies the effect, by increasing the height of the stability potential barrier and, consequently, the depth of the minimum of the graviton zero-mode on the brane.  Finally, we study the particular cases of $F(R)+T$ and $R+G\left(T\right)$, for which only the scalar fields $\varphi$ or $\psi$ are present, respectively, and we prove that the presence of the scalar field $\varphi$ is essential in the development of internal structure on the brane.
\end{abstract}
\maketitle

\section{Introduction}

While preserving the several positive results of Einstein's theory, extended theories of gravity can be considered as a new paradigm to address conceptual and experimental problems recently emerged in Astrophysics, Cosmology and High Energy Physics. Modifications of the Einstein's theory of gravity by replacing the curvature scalar $R$ in the action by the generic function $F(R)$ \cite{Capozziello:2011et,Nojiri:2010wj} or by including new scalars constructed using the Riemann tensor as the Gauss-Bonnet scalar $G=R ^2-4R_{\mu\nu}R^{\mu\nu} +R_{\mu\nu\alpha\beta} R^{\mu\nu\alpha\beta}$ \cite{Rippl:1995bg, Solodukhin:1994pc, Noh:1996da, Charmousis:2002rc} have been the subject of several recent studies. These proposals suggest viable routes to address important physical problems, e.g. cosmic inflation, dark energy, and local gravity constraints \cite{DeFelice:2010aj, Hwang:2001pu}. In addition to curvature scalars, proposals for modified theories of gravity have also considered the inclusion of matter terms, such as the so-called $f(T)$-gravity, where $T$ is the trace of the stress-energy tensor \cite{Harko:2011kv,Cai:2015emx}, or modifications to the way matter fields interact, called K-fields \cite{K1,K2,K3,K4,K5,Adam:2007ag}. The structuring of these ideas in different perspectives and models has been the subject of interesting studies over the last three decades.

In connection with superstring theories, which requires the presence of extra spatial dimensions \cite{Barton}, in the seminal work of Randall-Sundrum \cite{Randall:1999vf} an interesting model was introduced, dealing with a five-dimensional anti de-Sitter warped geometry engendering an extra spatial dimension of infinite extent. This possibility, which deals with a single brane in a factorized geometry, nicely reproduces four-dimensional gravity and is known as the thin brane scenario. Soon after, it was modified to include a source scalar field, giving rise to what is now known as the thick brane scenario \cite{Goldberger:1999uk,Skenderis:1999mm,DeWolfe:1999cp,Csaki:2000fc}. In the thick braneworld scenario, modified gravity has proven to be of major importance to extend the plethora of behaviors and results \cite{Afonso:2007gc,Bazeia:2013oha, Bazeia:2013uva, Bazeia:2015dna,Bazeia:2015owa,Correa:2015qma,Correa:2015ako, Rosa:2020uli}. In this perspective, important physical characteristics were analyzed in the context of theories such as $f(R)$, $f(R,T)$ and in mimetic gravity, with or without the inclusion of a Cuscuton term in the dynamics of the brane source fields \cite{Afonso:2007gc,Bazeia:2013oha,Bazeia:2013uva, Bazeia:2015dna,Bazeia:2015owa,Correa:2015qma,Correa:2015ako, Rosa:2020uli,Zhong,Guo:2018tpo,Bazeia:2021jok,Sui:2021uic}. In particular, in the presence of the Cuscuton term one usually investigates its contribution to change the internal structure of the thick brane. 

In the presence of the modifications to gravity described above, it is not unusual that the high degree of complexity of the system of equations prevents one to obtain analytical solutions in general cases. It is thus common to recur to numerical methods to solve this problem. In particular, braneworld models in the general $f\left(R,T\right)$ and particular $F\left(R\right)+T$ and $R+G\left(T\right)$ cases has been studied numerically \cite{Rosa:2021tei} in the scalar-tensor representation of the theory, where the two extra scalar degrees of freedom are considered explicitly via the introduction of two auxiliary scalar fields \cite{Rosa:2021teg}. In this work, we build upon these previous results \cite{Rosa:2021tei,Rosa:2021teg} by introducing a Cuscuton term for the dynamics of the source field of the brane, to see how it contributes to modify the geometry of the brane.

The Cuscuton dynamics are implemented though the addition of a scalar field without its own dynamical degrees of freedom; see, e.g., Refs. \cite{CG1,CG2}. In particular, it is shown that the Cuscuton action can model the continuum limit of the evolution of a field which is protected against quantum corrections at low energies. Moreover, in studies of cosmological bounce models based on Cuscuton gravity \cite{CC01,CC02}, the absence of ghosts or curvature instabilities was proven. One also finds that power-law inflation with the addition of Cuscuton dynamics can ameliorate inflation models and make predictions consistent with observations \cite{Ito:2019fie}. These are results of current interest, and have motivated the use of modified dynamics to describe Cuscuton kinks and branes \cite{Andrade:2018afh}. The fact that the Cuscuton dynamics appear to be more effective at low energies has inspired us to study how it can be used to change the geometry of the brane and, in the present work, we deal with this within the context of a five-dimensional braneworld model in the modified $f(R,T)$ scalar-tensor gravity, with a single extra spatial dimension of infinite extent. Our analysis includes also the analysis of the separable cases $F(R)+T$ and $R+G(T)$, which allow to verify the importance of the Cuscuton term on these two stable braneworld scenarios, with distinct geometry and matter contributions. 

In order to investigate these issues, we organize the present work as follows. In Sec.~\ref{sec01Stability} we introduce the general formalism that describes a $f(R,T)$-brane in a scalar-tensor representation with a Cuscuton term in the matter Lagrangian and we study two models numerically. In Sec.~ \ref{separableCases} we investigate the particular cases $F(R)+T$ and $R+G(T)$. In Sec.~ \ref{secStability} we perform the stability analysis of these models in the scalar-tensor representation of the theory, and finally in Sec.~ \ref{sec:conclusion} we trace our conclusions and present comments on future prospects of this work.

\section{Analysis of the general $f\left(R,T\right)$ case}\label{sec01Stability}

\subsection{Theory and field equations}

Let us consider the general $f\left(R,T\right)$ theory of gravity in $4+1$ dimensions, which is represented by an action $S$ of the form
\begin{equation}\label{actiongeo}
S=\frac{1}{2\kappa^2}\int_\Omega\sqrt{|g|}f\left(R,T\right)d^5x+S_m\left(g_{ab},\phi\right),
\end{equation}
where $\kappa^2=8\pi G_5/c^4$, where $G_5$ is the 5-dimensional gravitational constant and $c$ is the speed of light, $\Omega$ is a 5-dimensional spacetime manifold described by a set of coordinates $x^a$, $|g|$ is the absolute value of the determinant of the metric $g_{ab}$, $f$ is an arbitrary function of the Ricci scalar $R=g^{ab}R_{ab}$, where $R_{ab}$ is the Ricci tensor of the metric $g_{ab}$, and the trace $T=g^{ab}T_{ab}$ of the stress-energy tensor $T_{ab}$, $S_m$ is the matter action defined by $S_m=-\int\mathcal L_s d^5x$, where $\mathcal L_s$ is the Lagrangian density of the source of the braneworld model, considered minimally coupled to the metric $g_{ab}$, and $\phi$ denotes the source field.

One can derive the field equations of the theory via the variation of Eq.~~\eqref{actiongeo} with respect to the metric $g_{ab}$, leading to
\begin{eqnarray}
\frac{\partial f}{\partial R}R_{ab}&-&\frac{1}{2}f\left(R,T\right)g_{ab}-\left(\nabla_a\nabla_b-g_{ab}\Box\right)\frac{\partial f}{\partial R}
	\nonumber \\
&=&\kappa^2 T_{ab}-\frac{\partial f}{\partial T}\left(T_{ab}+\Theta_{ab}\right),\label{fieldgeo}
\end{eqnarray}
where $\nabla_a$ represents covariant derivatives and $\Box\equiv\nabla^a\nabla_a$ the d'Alembert operator, both written with respect to the metric $g_{ab}$. The stress-energy tensor $T_{ab}$ is defined in the usual way as the variation of the source Lagrangian density $\mathcal L_s$ with respect to the metric $g_{ab}$, i.e.,
\begin{equation}\label{defTab}
T_{ab}=-\frac{2}{\sqrt{|g|}}\frac{\delta\left(\sqrt{|g|}\mathcal L_s\right)}{\delta g^{ab}},
\end{equation}
and $\Theta_{ab}$ is in turn defined in terms of the variation of the stress-energy tensor $T_{ab}$ with respect to the metric $g_{ab}$, i.e.,
\begin{equation}\label{deftheta}
\Theta_{ab}=g^{cd}\frac{\delta T_{cd}}{\delta g^{ab}}.
\end{equation}
Once the form of either the stress-energy tensor $T_{ab}$ or the source Lagrangian density $\mathcal L_s$ are given explicitly, one can obtain the explicit form of the tensor $\Theta_{ab}$.

Similarly to what happens in the $f\left(R\right)$ theory of gravity, one can derive a dynamically equivalent scalar-tensor representation of the theory by introducing two auxiliary scalar fields $\alpha$ and $\beta$. To do so, let us rewrite Eq.~~\eqref{actiongeo} in the form
\begin{eqnarray}
S=\frac{1}{2\kappa^2}\int_\Omega\!\sqrt{|g|}\left[f\left(\alpha,\beta\right)+\frac{\partial f}{\partial \alpha}\left(R-\alpha\right)+\right.
	\nonumber\\
\left.+\frac{\partial f}{\partial\beta}\left(T-\beta\right)\right]d^5x+S_m\left(g_{ab},\phi\right).\label{auxaction1}
\end{eqnarray}
Equation \eqref{auxaction1} depends on three independent quantities, the metric $g_{ab}$ and the auxiliary scalar fields $\alpha$ and $\beta$, plus any source field that might be present. The equations of motion for these auxiliary fields obtained via the variation of Eq.~\eqref{auxaction1} are
\begin{equation}\label{auxeom1}
f_{\alpha\alpha}\left(R-\alpha\right)+f_{\alpha\beta}\left(T-\beta\right)=0,
\end{equation}
\begin{equation}\label{auxeom2}
f_{\beta\alpha}\left(R-\alpha\right)+f_{\beta\beta}\left(T-\beta\right)=0,
\end{equation}
respectively, where the subscripts $\alpha$ and $\beta$ denote partial derivatives with respect to these fields, and $f_{\alpha\beta}=f_{\beta\alpha}$ for any well-behaved function $f\left(\alpha,\beta\right)$ satisfying the Schwartz theorem. Equations \eqref{auxeom1} and \eqref{auxeom2} can be rewritten as a matrix equation $\mathcal M \textbf{X}=0$ in the form
\begin{equation}\label{matrixeq}
\mathcal M \textbf{X}=\begin{pmatrix}
f_{\alpha\alpha} & f_{\alpha\beta} \\
f_{\beta\alpha} & f_{\beta\beta}
\end{pmatrix}
\begin{pmatrix}
R-\alpha\\
T-\beta
\end{pmatrix}
=0.
\end{equation}
The system in Eq.~\eqref{matrixeq} will present a unique solution if and only if the determinant of the matrix $\mathcal M$ does not vanish, which imposes a constraint on the second-order partial derivatives of $f\left(\alpha,\beta\right)$ of the form $f_{\alpha\alpha}f_{\beta\beta}\neq f_{\alpha\beta}^2$. Provided that this condition is valid, the unique solution of Eqs.\eqref{auxeom1} and \eqref{auxeom2} is $\alpha=R$ and $\beta=T$. Inserting these results back into Eq.~\eqref{auxaction1} one verifies that the original action in Eq.~\eqref{actiongeo} is recovered and the two representations are dynamically equivalent.

One can now define two scalar fields $\varphi$ and $\psi$, along with a scalar potential $U\left(\varphi,\psi\right)$ as
\begin{equation}\label{defscalar}
\varphi=\frac{\partial f}{\partial R}, \qquad \psi=\frac{\partial f}{\partial T},
\end{equation}
\begin{equation}\label{defpotential}
U\left(\varphi,\psi\right)=\varphi R+\psi T-f\left(R,T\right),
\end{equation}
in such a way that the auxiliary action in Eq.~ \eqref{auxaction1} can be rewritten in the scalar-tensor representation as
\begin{eqnarray}
S\!=\!\frac{1}{2\kappa^2}\!\int_\Omega\!\!\sqrt{|g|}\,\big[\varphi R+\psi T-U\left(\varphi,\psi\right)\big]d^5x
+S_m(g_{ab},\phi).\nn\,\,\label{actionst}
\end{eqnarray}
Again, Eq.~ \eqref{actionst} depends on three independent quantities, in this case the metric $g_{ab}$ and the two scalar fields $\varphi$ and $\psi$, plus any source field. Varying Eq.~\eqref{actionst} with respect to these variables respectively yields
\begin{eqnarray}
&&\varphi R_{ab}-\frac{1}{2} g_{ab}\left(\varphi R+\psi T-U\right)
-\left(\nabla_a\nabla_b-g_{ab}\Box\right)\varphi
 \nonumber \\
&&\qquad \qquad \qquad =\kappa^2 T_{ab}-\psi\left(T_{ab}+\Theta_{ab}\right),
\label{fieldst}
\end{eqnarray}
\begin{equation}\label{eomphi}
U_\varphi=R,
\end{equation}
\begin{equation}\label{eompsi}
U_\psi=T,
\end{equation}
where subscripts $\varphi$ and $\psi$ denote partial derivatives with respect to these fields, respectively. As the numerical analysis of the standard dynamics in the scalar tensor representation of the general $f\left(R,T\right)$ gravity was already studied in a previous publication \cite{Rosa:2021tei}, in this work we will focus solely in the Cuscuton dynamics and their consequences.

\subsection{Cuscuton dynamics}

Let us consider the source matter to be described by a single dynamical scalar field $\phi$ with an interaction potential $V\left(\phi\right)$ and a Cuscuton term controlled by a positive real parameter $\alpha$. The source Lagrangian density that describes this distribution of matter is given by
\begin{equation}\label{actionchi}
\mathcal L_s=\frac{1}{2}\nabla_c\phi\nabla^c\phi+\alpha\sqrt{|\nabla_c\phi\nabla^c\phi|}-V\left(\phi\right)\,.
\end{equation}
To simplify the notation in the upcoming calculations, let us define a quantity $X$ as
\begin{equation}
X=\alpha\frac{\sqrt{|\nabla_c\phi\nabla^c\phi|}}{\nabla_c\phi\nabla^c\phi}.
\end{equation}
Inserting Eq.~\eqref{actionchi} into the definition of the stress-energy tensor $T_{ab}$ in Eq.~\eqref{defTab} yields
\begin{equation}\label{tmn}
T_{ab}=\left(1+X\right)\nabla_a\phi\nabla_b\phi-g_{ab}\mathcal L_s.
\end{equation}
The explicit form of $T_{ab}$ in Eq.~\eqref{tmn} allows one to obtain the tensor $\Theta_{ab}$ from Eq.~\eqref{deftheta} via the variation with respect to $g^{ab}$. The result is
\begin{equation}\label{theta}
\Theta_{ab}=-\left(\frac{5}{2}+3X\right)\nabla_a\phi\nabla_b\phi-g_{ab}\mathcal L_s\,.
\end{equation}

Finally, the equation of motion for the field $\phi$ can be obtained by taking a variation of the matter action with respect to $\phi$ using the source Lagrangian $\mathcal L_s$ in Eq.~\eqref{actionchi}. Note that given the dependence in $T$ of the gravitational sector of the action, the equation of motion for $\phi$ will also feature contributions from this sector. The resultant equation is
\begin{eqnarray}\label{eomchi}
&&\left[1+X+\frac{\psi}{2\kappa^2}\left(3+4X\right)\right]\Box\phi+\left(\frac{4\psi}{2\kappa^2}+1\right)\nabla_c\phi\nabla^cX+ \nonumber \\
&&+\frac{1}{2\kappa^2}\left(3+4X\right)\nabla_c\phi\nabla^c\psi+\left(\frac{5\psi}{2\kappa^2}+1\right)V_\phi=0\,,
\end{eqnarray}
where the subscript $\phi$ denotes a derivative with respect to that field.

\subsection{Equations and solutions}

Let us now consider the standard metric for the braneworld model with an additional dimension $y$ described by a line element given by
\begin{equation}\label{metricbrane}
    ds^2=e^{2A}\eta_{\mu\nu}dx^\mu dx^\nu-dy^2\,,
\end{equation}
where $\eta_{\mu\nu}$ is the four-dimensional Minkowski metric with signature $(+\,-\,-\,-)$, and greek indices run from $0$ to $3$. Furthermore, we assume the system to be static, i.e., all quantities considered are functions solely of the extra dimension. In this form, the warp function, auxiliary fields, and the source field of the brane are written as $A=A(y)$, $\psi=\psi(y)$, $\varphi=\varphi(y)$, and $\phi=\phi(y)$, respectively. Furthermore, in the following calculations we will consider a system of geometrized units for which $\kappa^2=2$. Given the homogeneity and isotropy of the metric in Eq.~\eqref{metricbrane} in the four-dimensional hypersurface, the field equations in Eq.~\eqref{fieldst} feature only two linearly independent components, namely
\begin{eqnarray}
&&3\varphi\left( A''+2A'^2\right)+3\varphi' A'+\varphi''+\frac{1}{2}U=\label{field1}\\
&=&-\left(\frac{5}{2}\psi+2\right)V-\left(\frac{3}{4}\psi+1\right)\phi'^2-2\alpha\phi'\left(1+\psi\right),\nonumber
\end{eqnarray}
\begin{eqnarray}
&&6\varphi A'^2+4\varphi' A'+\frac{1}{2}U=\label{field2}\\
&=&-\left(\frac{5}{2}\psi+2\right)V+\left(\frac{3}{4}\psi+1\right)\phi'^2\nonumber.
\end{eqnarray}
These two equations can be linearly combined to obtain the simpler relation
\begin{equation}\label{fieldsum}
3\varphi A''+\varphi''-\varphi'A'=-\left(\frac{3}{2}\psi+2\right)\phi'^2-2\alpha\phi'\left(1+\psi\right).
\end{equation}
The equations of motion for the auxiliary scalar fields $\varphi$ and $\psi$ from Eqs.\eqref{eomphi} and \eqref{eompsi} become
\begin{equation}\label{kgphi}
U_\varphi=8A''+20A'^2,
\end{equation}
\begin{equation}\label{kgpsi}
U_\psi=\frac{3}{2}\phi'^2+5V-4\alpha\phi',
\end{equation}
and finally the equation of motion for the source field $\phi$ given in Eq.~\eqref{eomchi} yields
\begin{eqnarray}
&&\left(\frac{3}{4}\psi+1\right)\phi''+\left(\frac{3}{4}\psi'+\frac{3}{2}\psi A'+4A'\right)\phi'+\nonumber\\
&&+\alpha\left(4A'+4\psi A'+\psi'\right)=\left(\frac{5}{4}\psi+1\right)V_\phi.\label{kgchi}
\end{eqnarray}
Note that Eqs.\eqref{kgphi} to \eqref{kgchi} depend on partial derivatives of the potential functions, i.e., $U_\varphi$, $U_\psi$ and $V_\phi$. These partial derivatives can be written in terms of derivatives of $U$, $V$, $\varphi$, $\psi$ and $\phi$ with respect to $y$ via the use of the chain rule. In particular, for the potential $V$ which is a function solely of $\phi$, one can write $V'=V_\phi \phi'$, and use this relation to eliminate $V_\phi$ from Eq.~\eqref{kgchi}. For the potential $U$ however, since it is a function of both $\varphi$ and $\psi$, the derivatives $U_\varphi$ and $U_\psi$ can not be directly eliminated from the equations, but a relationship between them can be found via the same method, which yields
\begin{equation}\label{Urel}
U'=U_\varphi \varphi'+U_\psi\psi'.
\end{equation}
As a result, the system of Eqs.\eqref{field1}, \eqref{field2}, \eqref{kgphi}, \eqref{kgpsi}, \eqref{kgchi}, and \eqref{Urel} form a system of six equations, from which only five are linearly independent. This can be proved by taking the derivative of Eq.~\eqref{field2} and using the system equations to cancel the quantities $A''$, $\phi''$, $\varphi''$, $V$, $U$, and $U'$. The result is an identity, thus proving that these equations are linearly dependent, and supporting that one of these equations may be discarded without loss of generality. Due to the complicated nature of the field equations, we chose to replace Eqs.\eqref{field1} and \eqref{field2} by Eq.~\eqref{fieldsum}. Finally, one can use Eqs.\eqref{kgphi} and \eqref{kgpsi} to eliminate the quantities $U_\varphi$ and $U_\psi$ from Eq.~\eqref{Urel} (for a detailed proof of why $U_\varphi$, $U_\psi$ and $U$ can be considered all independent quantities, we refer the reader to Sec.~III B of \cite{Rosa:2021tei}), thus obtaining
\begin{equation}\label{Urelfinal}
U'=\left(8A''+20A'^2\right)\varphi'+\left(\frac{3}{2}\phi'^2+5V-4\alpha\phi'\right)\psi'\,.
\end{equation}

We are thus left with a system of three independent equations, Eqs.\eqref{fieldsum}, \eqref{kgchi}, and Eq.~\eqref{Urelfinal}, for the six independent quantities $A$, $\varphi$, $\psi$, $\phi$, $V$, and $U$. This implies that the system is underdetermined and one can impose three constraints to close the system. In the following we pursue such an analysis by choosing explicit forms for $A$, $\phi$ and $V$, and leaving the auxiliary fields and potential as unknown quantities. Given the complicated nature of the equations, analytical solutions are not attainable in this general case, and we shall focus on numerical methods.

\subsubsection{Model 1}\label{sec:numerical1}

As a first example, let us impose the following constraints on the quantities $A$, $\phi$ and $V$,
\begin{equation}\label{A1}
A\left(y\right)=A_0\log\big[\sech\left(k y\right)\big]+A_1\tanh\left(ky\right),
\end{equation}
\begin{equation}\label{phi1}
\phi\left(y\right)=\phi_0\tanh\left(ky\right),
\end{equation}
\begin{equation}\label{V1}
V\left(\phi\right)=\frac{1}{2}\big(W_\phi+\alpha\big)^2-\frac{4}{3}W\left(\phi\right)^2,
\end{equation}
where the parameters $A_0>0$, $A_1<0$, $k$, and $\phi_0$ are real constants, and the function $W\left(\phi\right)$, known as the super-potential, was defined as
\begin{equation}\label{W1}
W\left(\phi\right)=\phi-\frac{1}{3}\phi^3.
\end{equation}
As shown in Ref. \cite{Bazeia:2021jok}, if one uses $W$ as given by the above Eq. \eqref{W1}, the set of Eqs. \eqref{A1} -- \eqref{V1} give the solutions of the model in the presence of the Cuscuton term for the standard gravity, that is, for $F=R$. This is important, since it proves that the generalized theory preserves the desired qualitative behavior present in the standard braneworld scenario. For instance, it ensures that the energy of the brane, which is obtained from $E=\int \rho dy$, where $\rho=T_{00}=-e^{2A}{\cal L}_s$, has the usual profile, leading to the thin brane limit and ensuring the asymptotic limit of the model, such that $\Lambda_5=V(\pm\phi_0)<0$, which makes the bulk asymptically $AdS_5$. In this sense, in order to avoid undesired issues, we then follow the direction described before for the standard gravity with the Cuscuton term, taking the corresponding relations as constraints to guide the numerical investigation to be done below.

\begin{figure}
\includegraphics[scale=0.8]{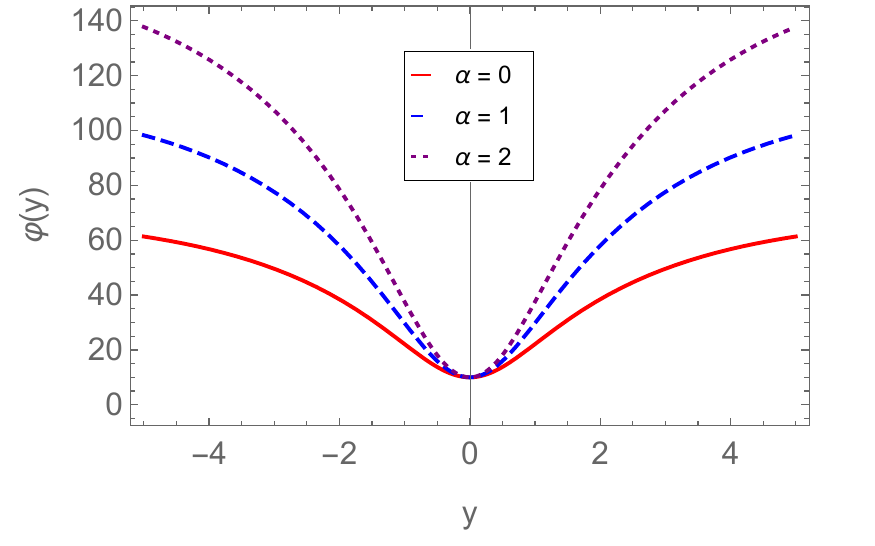}
\includegraphics[scale=0.8]{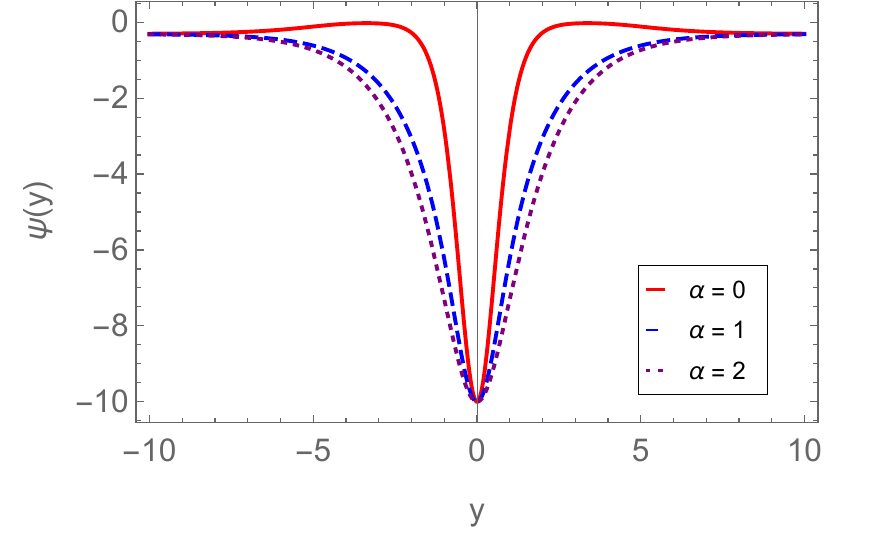}
\includegraphics[scale=0.8]{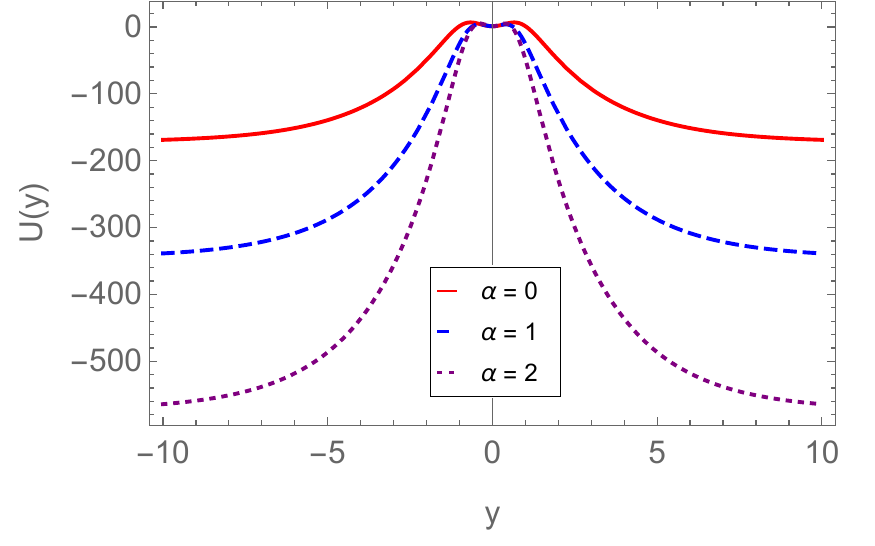}
\caption{Numerical solutions for $\varphi(y)$ (top panel), $\psi(y)$ (middle panel) and $U(y)$ (bottom panel) from Eqs. \eqref{fieldsum}, \eqref{kgchi}, and Eq.~\eqref{Urelfinal} with $\varphi_0=10=-\psi_0$, $U_0=k=\phi_0=1$, $A_0=4/9$, and $A_1=-1/9$, for different values of $\alpha$.}
\label{fig:numerical1}
\end{figure}

To solve the system of Eqs. \eqref{fieldsum}, \eqref{kgchi}, and Eq.~\eqref{Urelfinal} numerically, one has to impose four boundary conditions on the brane, i.e., at $y=0$, namely $\varphi\left(y=0\right)=\varphi_0$, $\psi\left(y=0\right)=\psi_0$, $U\left(y=0\right)=U_0$, and $\varphi'\left(y=0\right)=0$, the latter guaranteeing that the solutions are even. Since in a previous work (see \cite{Rosa:2021tei}) we have proven that the brane only develops internal structure if the signs of $\varphi_0$ and $\psi_0$ are contrary, and in this work we are interested in studying the effects of the Cuscuton term in the internal structure of the brane, we shall deal solely with boundary conditions satisfying this requirement, e.g., $\varphi_0=10=-\psi_0$. The solutions for $\varphi(y)$, $\psi(y)$ and $U(y)$ are plotted in Fig.~\ref{fig:numerical1}. Although the general behavior of the fields is not massively affected, it is clear that a variation in $\alpha$ has a non-negligible effect in the shape of both the fields and the potential.

\subsubsection{Model 2}\label{sec:numerical2}

\begin{figure}
\includegraphics[scale=0.88]{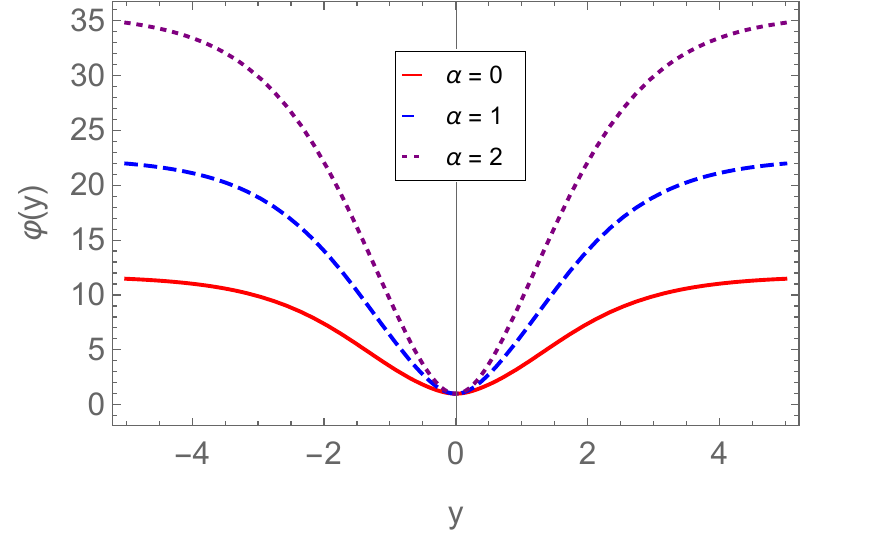}
\includegraphics[scale=0.9]{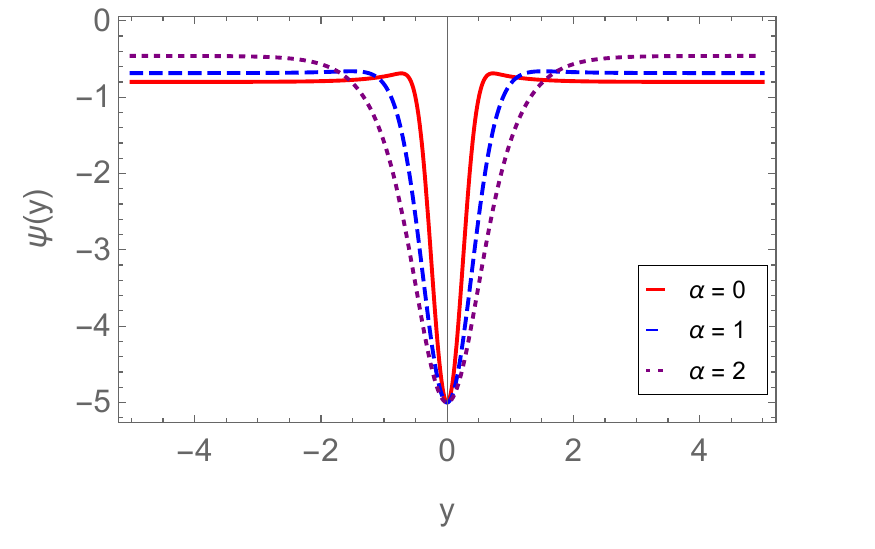}
\includegraphics[scale=0.84]{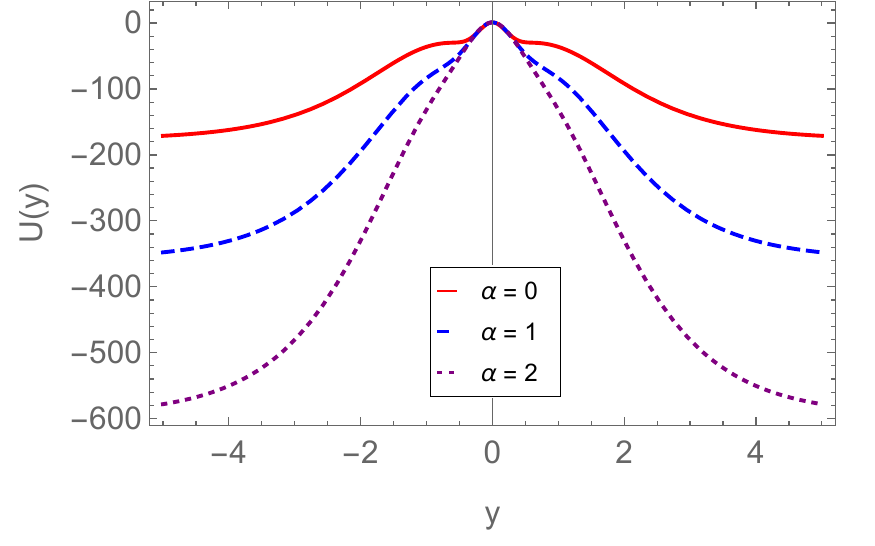}
\caption{Numerical solutions for $\varphi(y)$ (top panel), $\psi(y)$ (middle panel) and $U(y)$ (bottom panel) from Eqs. \eqref{fieldsum}, \eqref{kgchi}, and Eq.~\eqref{Urelfinal} with $\varphi_0=U_0=k=\phi_0=1$, $\psi_0=-5$, $A_0=4/9$, and $A_1=-1/9$, for different values of $\alpha$.}
\label{fig:numerical2}
\end{figure}

In this second example, we impose the following constraints on the quantities $A$, $\phi$ and $V$,
\begin{equation}\label{A2}
A\left(y\right)=A_0\ln\left[\sech\left(k y\right)\right],
\end{equation}
\begin{equation}\label{phi2}
\phi\left(y\right)=\phi_0\arctan\left[\tanh\left(ky\right)\right],
\end{equation}
\begin{equation}\label{V2}
V\left(\phi\right)=\frac{1}{2}\big(W_\phi+\alpha\big)^2-\frac{4}{3}W\left(\phi\right)^2,
\end{equation}
where the parameters $A_0>0$, $k$, and $\phi_0$ are again constants, and the superpotential $W\left(\phi\right)$, in this case is defined as
\begin{equation}\label{W2}
W\left(\phi\right)=\sin\phi.
\end{equation}
Similarly to the previous model, we see Eqs. \eqref{A2} -- \eqref{V2} as constraints that appear from the study of brane in the case of standard gravity with the Cuscuton, using $W(\phi)$ as in Eq. \eqref{W2}. Again, we introduce the constraints in order to avoid undesired changes in the model in the scalar-tensor representation of the generalized theory. Here we have to impose four boundary conditions on the brane at $y=0$, namely $\varphi\left(y=0\right)=\varphi_0$, $\psi\left(y=0\right)=\psi_0$, $U\left(y=0\right)=U_0$, and $\varphi'\left(y=0\right)=0$, to be able to integrate the system of Eqs. \eqref{fieldsum}, \eqref{kgchi}, and Eq.~\eqref{Urelfinal} numerically. Again, to study the effects of the Cuscuton term on the internal structure of the brane, we choose $\varphi_0=1$, $\psi_0=-5$. The solutions for $\varphi(y)$, $\psi(y)$ and $U(y)$ are plotted in Fig.~\ref{fig:numerical2}. Similarly to model 1, a variation in the Cuscuton parameter $\alpha$ affects the shapes of the auxiliary fields and potential in a non-negligible matter, although the general behaviors are not altered.

We noticed that in the two models investigated above, the Cuscuton parameter modifies the asymptotic behavior of the field $\varphi$, but it only changes the field $\psi$ near the origin. However, these two effects combine when one investigates the potential $U$.

\section{Analysis of particular separable cases}\label{separableCases}

In a previous work \cite{Rosa:2021tei}, it was shown that in $f\left(R,T\right)$ gravity with a standard scalar-field source, the brane only develops an internal structure when the two extra scalar degrees of freedom are present, i.e., when an arbitrary dependence on both $R$ and $T$ is present in the action and, consequently, the to scalar fields $\varphi$ and $\psi$ appear in the scalar-tensor representation of the theory. In the particular cases for which one of these dependencies is suppressed, e.g. $f\left(R,T\right)=F\left(R\right)+T$ or $f\left(R,T\right)=R+G(T)$, and thus only the scalar field $\varphi$ or $\psi$ appear in the scalar-tensor representation, respectively, no combinations of the parameters tested was capable of inducing internal structure on the brane. This conclusion was in agreement with the results previously obtained from a Bloch brane, where internal structure was induced by a combination of two scalar fields \cite{Bazeia:2021jok}.

However, it was also shown in Ref.\cite{Bazeia:2021jok} that if the brane is sourced by a scalar field with Cuscuton dynamics, internal structure can be induced even in the presence of a single extra scalar degree of freedom. It is thus worth studying how the Cuscuton scalar field $\phi$ interacts separately with the scalar fields $\varphi$ and $\psi$ in these separable particular cases in the scalar-tensor representation, in which situations it is possible to induce an internal structure on the brane, and what is the effect of the Cuscuton term in the stability potentials and structure of the graviton zero mode. We now turn to this.

\subsection{Case of $f\left(R,T\right)=F\left(R\right)+T$}\label{sec:spec1}

Let us start by considering the particular case for which $f\left(R,T\right)=F\left(R\right)+T$. The general method described in the previous section is no longer valid in this situation since $f_{RT}=f_{TT}=0$, and thus $\det \mathcal M=0$. Following Ref.\cite{Rosa:2021tei}, it was shown that the scalar-tensor representation associated with such a particular case is described by an action $S$ of the form
\begin{equation}\label{spec1_action}
S=\frac{1}{2\kappa^2}\int_\Omega\sqrt{|g|}\left[\varphi R+T-U\left(\varphi\right)\right]d^5x+S_m\left(g_{ab},\phi\right).
\end{equation}
Notice how the scalar field $\psi$ is completely absent from this action, both explicitly and via the potential $U$, which is now a function of $\varphi$ only. There are thus only two equations of motion to be derived from the variational method with respect to $g_{ab}$ and $\psi$, which are
\begin{eqnarray}\label{spec1_field}
&&\varphi R_{ab}-\frac{1}{2}g_{ab}\left(\varphi R+T-U\right)-\\
&&-\left(\nabla_a\nabla_b-g_{ab}\Box\right)\varphi=\kappa^2T_{ab}-\left(T_{ab}+\Theta_{ab}\right),\nonumber
\end{eqnarray}
\begin{equation}\label{spec1_kgphi}
U_\varphi=R.
\end{equation}
Notice how these equations of motion could be obtained from the general case by imposing $U\left(\varphi,\psi\right)=U\left(\varphi\right)$ and $\psi=1$, see Eqs.\eqref{fieldst} and \eqref{eomphi}, but they could not be obtained as a limit of the general system because Eq.~\eqref{eompsi} would force $T=0$, whereas here the distribution of matter is still arbitrary. The complete set of independent equations for this case, considering a metric defined by Eq.~\eqref{metricbrane} and a distribution of matter of the form of Eq.~\eqref{actionchi}, becomes
\begin{equation}\label{spec1_fieldsimp}
3\varphi A''+\varphi''-\varphi' A'=-\frac{7}{2}\phi'^2-4\alpha\phi',
\end{equation}
\begin{equation}\label{spec1_eqU}
U'=\big(8 A''+20 A'^2\big)\varphi',
\end{equation}
\begin{equation}\label{spec1_eqA}
\frac{7}{4}\phi''+\frac{11}{2}A'\phi'+8\alpha A'=\frac{9}{4}V_\phi,
\end{equation}
where Eq.~\eqref{spec1_fieldsimp} was obtained by subtracting the $(y,y)$ component from the $(t,t)$ component of Eq.~\eqref{spec1_field}, and Eq.~\eqref{spec1_eqU} via the use of the chain rule on  Eq.~ \eqref{spec1_kgphi}. It is interesting to note that even though we have reduced both the number of equations and variables by one in comparison to the general system (the equation of motion for $\psi$ and the field $\psi$ itself have been discarded), in this particular case we have one fewer degree of freedom to impose constraints in comparison to the general case. This difference arises from the fact that the potential $U$, being a function of solely $\varphi$, carries a single degree of freedom, whereas in the general case the potential $U\left(\varphi,\psi\right)$ carries two degrees of freedom. Thus, only two extra constraints can be imposed to close the system, e.g. the explicit form of $\phi$ and $V$, and $A$ must be computed from the equations. In this case, we impose
\begin{equation}\label{spec1_ansphi}
\phi\left(y\right)=\phi_0\tanh\left(ky\right),
\end{equation}
\begin{equation}\label{spec1_ansV}
V\left(\phi\right)=\frac{1}{2}\left(W_\phi+\alpha\right)^2-\frac{4}{3}W\left(\phi\right)^2,
\end{equation}
\begin{equation}\label{spec1_ansW}
W\left(\phi\right)=\phi-\frac{1}{3}\phi^3.
\end{equation}

\begin{figure}
\includegraphics[scale=0.8]{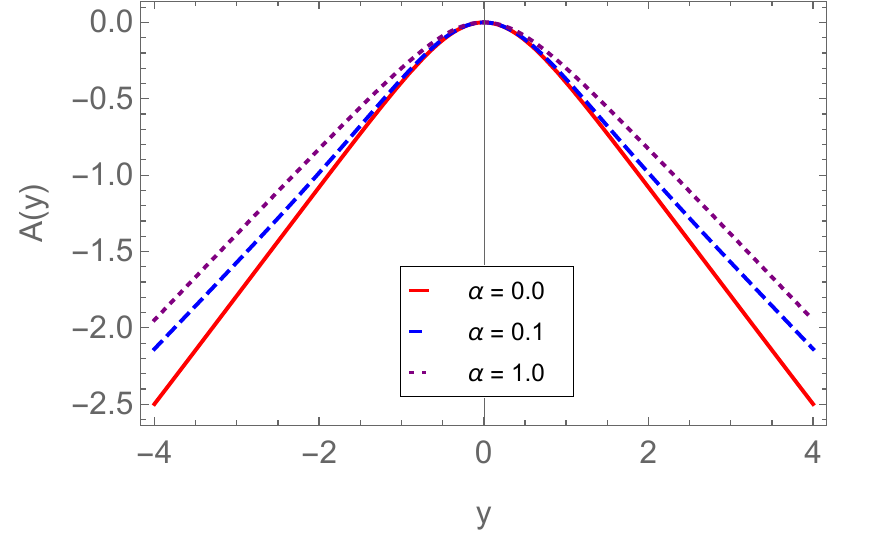}
\caption{Warp function $A\left(y\right)$ resulting from Eq.~\eqref{spec1_eqA} with $\phi_0=k=1$ for different values of $\alpha$.}
\label{fig:spec1_warp}
\end{figure}

Inserting Eqs.\eqref{spec1_ansphi} to \eqref{spec1_ansW} into the system of Eqs.\eqref{spec1_fieldsimp} to \eqref{spec1_eqA}, one verifies that Eq.~\eqref{spec1_eqA} becomes a decoupled differential equation for $A$ which can be directly integrated to find an analytical solution for $A$. We do not show this solution explicitly due to its size. Instead, we plot the warp function $A\left(y\right)$ in Fig.~\ref{fig:spec1_warp} for different values of $\alpha$.

\begin{figure}
\includegraphics[scale=0.55]{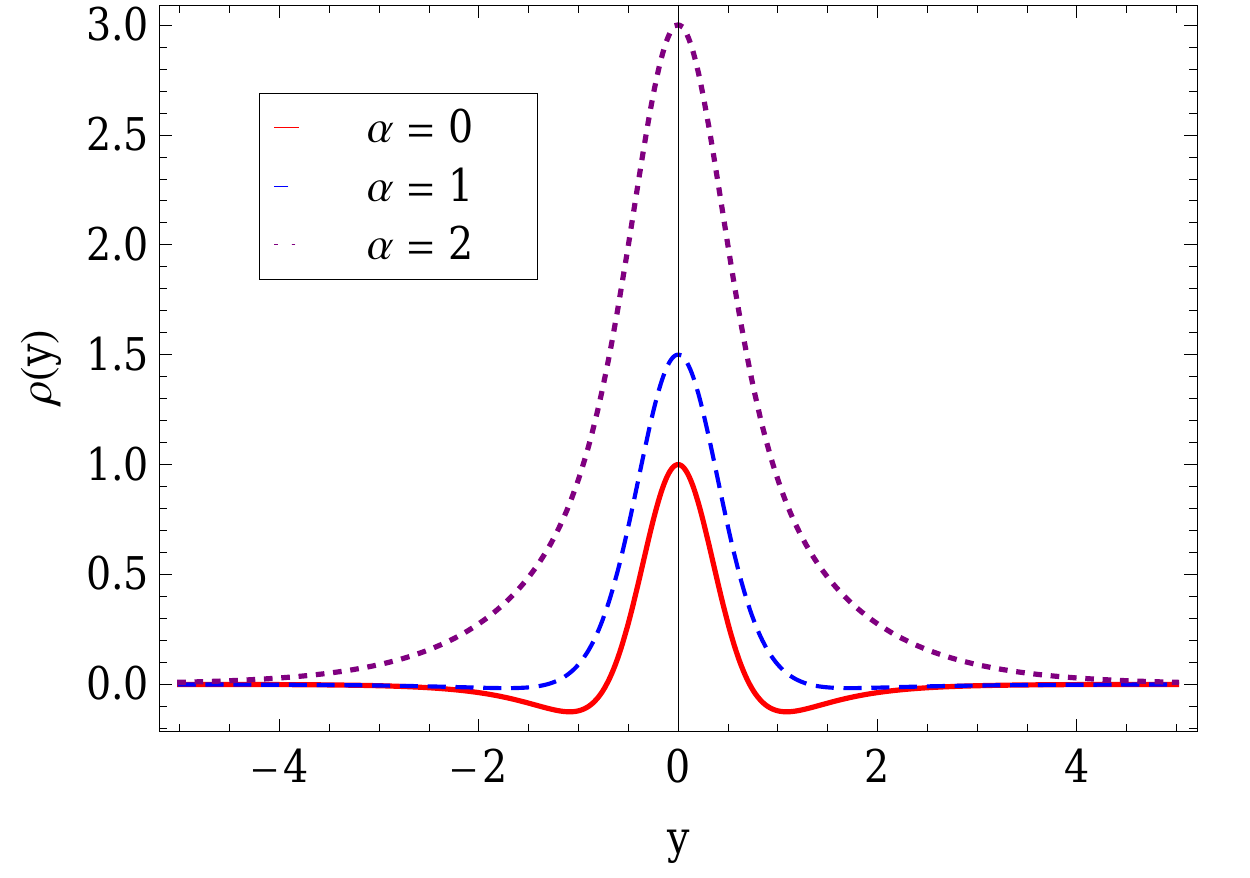}
\caption{Energy density $\rho(y)$ with $\phi_0=k=1$ for different values of $\alpha$.}
\label{fig:spec1_rho}
\end{figure}

Different from the models studied in the previous section, here the constraints introduced by Eqs. \eqref{spec1_ansphi} -- \eqref{spec1_ansW} can alter the energy density of the brane, since the warp function is now to be found from Eq. \eqref{spec1_eqA}. However, the profile of the energy density is qualitatively the same. To show this explicitly, we can construct the energy density from the expression 
$$
\rho=e^{2A} \Big(\frac12\phi^{\prime 2}-\alpha\phi'+V\Big).\label{energyDensity}
$$
We can use Eqs.\eqref{spec1_ansphi} -- \eqref{spec1_ansW} and the solution of Eq.\eqref{spec1_eqA} to get an analytical expression for the energy density. However, instead of writing the analytical expression, we decided to depict its shape in Fig. \ref{fig:spec1_rho}, for some values of $\alpha$.
On the other hand, the Eqs. \eqref{spec1_ansphi} -- \eqref{spec1_ansW} ensure the asymptotic behavior of the model, such that $\Lambda_5=V(\pm\phi_0)<0$, keeping the bulk asymptotically $AdS_5$. 

\begin{figure}
\includegraphics[scale=0.8]{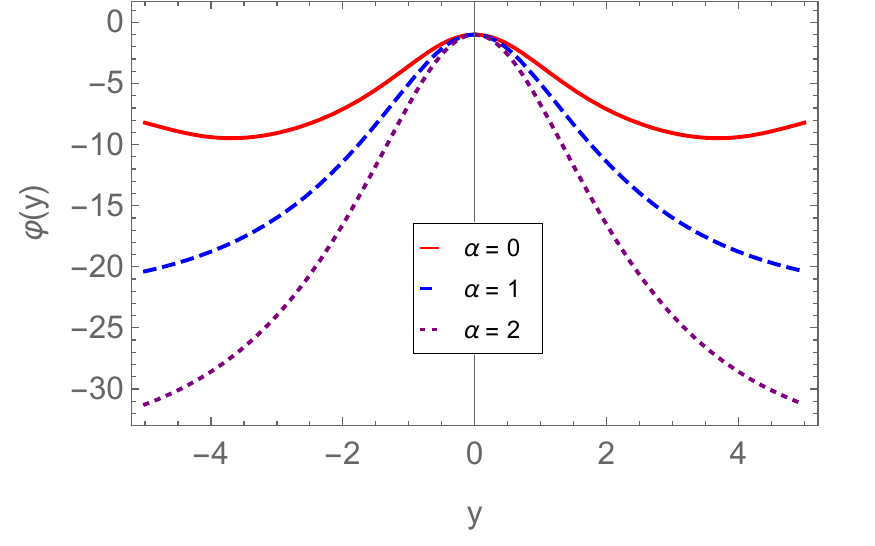}
\includegraphics[scale=0.8]{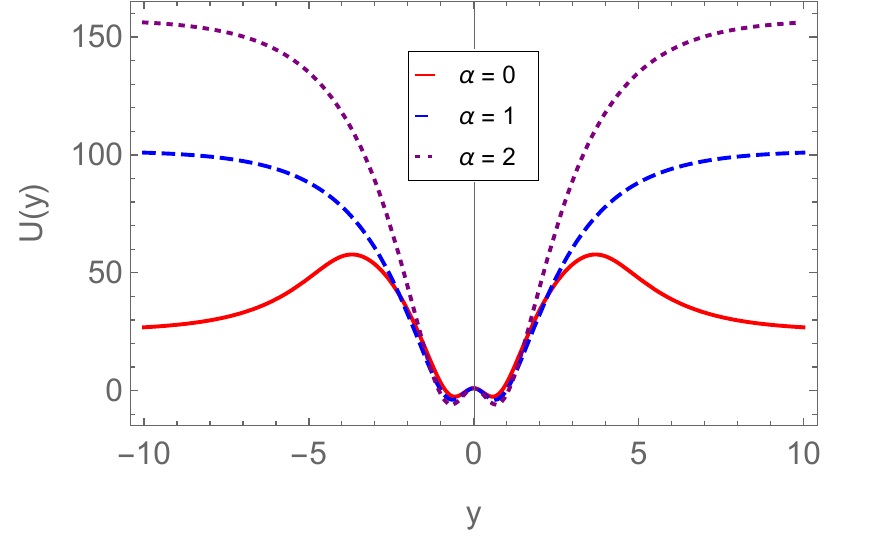}
\caption{Numerical solutions for $\varphi(y)$ (top panel) and $U(y)$ (bottom panel) from Eqs. \eqref{spec1_fieldsimp} and \eqref{spec1_eqU} with $\varphi_0=-1$, $U_0=k=\phi_0=1$, for different values of $\alpha$.}
\label{fig:spec1_sols}
\end{figure}

The solutions for $\varphi$ and $U$ must then be obtained numerically from the remaining equations, i.e., Eqs.\eqref{spec1_fieldsimp} and \eqref{spec1_eqU}, subjected to a set of boundary conditions $\varphi\left(y=0\right)=\varphi_0$, $U\left(y=0\right)=U_0$, and $\varphi'\left(y=0\right)=0$. In Fig.~\ref{fig:spec1_sols} we plot the solutions for $\varphi$ and $U$ under the boundary condition $\varphi_0=-1=-U_0$ and for different values of the parameter $\alpha$. It can be seen that for this combination of parameters the Cuscuton parameter $\alpha$ affects the shapes of $\varphi$ and $U$ and that these modifications can alter the qualitative behavior of these functions, unlike in the general case. For example, an increase in $\alpha$ can transform the potential $U$ from a double potential barrier into a double potential well, which simultaneously leads to a scalar field $\varphi$ that is monotonically decreasing away from the brane instead of achieving two minima.

\begin{figure}
\includegraphics[scale=0.8]{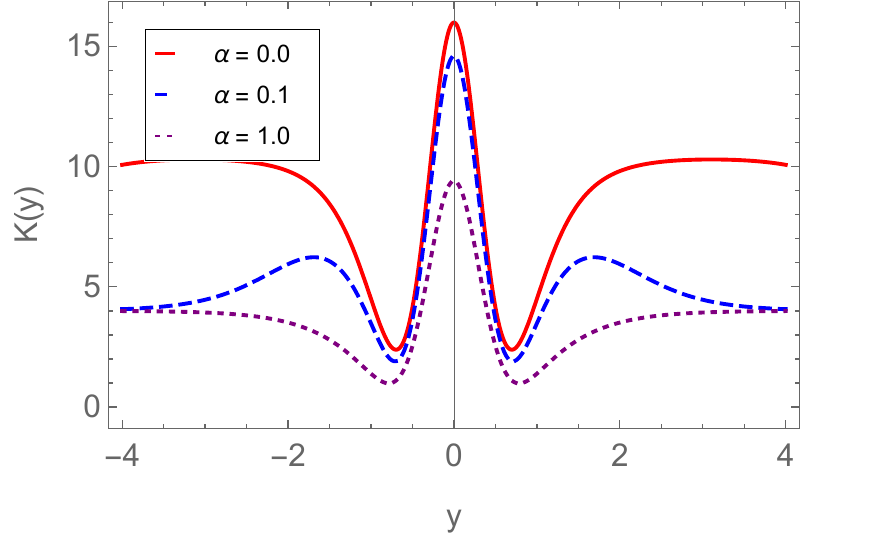}
    \caption{Numerical solutions for the Kretschmann scalar $K\left(y\right)$ from Eq.~\eqref{defKS} with $\phi_0=k=1$ for different values of $\alpha$.}
    \label{fig:kretschmann1}
\end{figure}

Since in this section the warp function $A$ was not the standard one but instead was obtained as a solution of Eq.~\eqref{spec1_eqA}, it is useful to compute the Kretschmann scalar to verify that no unwanted divergences appear in the system. The Kretschmann scalar is given by
\begin{equation}\label{defKS}
    K=40A^{\prime 4}+16A^{\prime\prime 2}+32A^{\prime 2}A^{\prime\prime}\,.
\end{equation}
The solutions for the Kretschmann scalar for the three cases studies in this section are plotted in Fig.~\ref{fig:kretschmann1}. It can be seen that it behaves appropriately in the presence of the Cuscuton term, for the values of the parameters used in this situation.

\subsection{Case of $f\left(R,T\right)=R+G\left(T\right)$}\label{sec:spec2}

Let us now turn to another particular case given by $f\left(R,T\right)=R+G\left(T\right)$. The general method is again not valid in this situation  since $f_{RR}=f_{RT}=0$, and thus $\det \mathcal M=0$. It was also shown in Ref.\cite{Rosa:2021tei} that the scalar-tensor representation associated with this particular case is described by the following action $S$:
\begin{equation}\label{spec2_action}
S=\frac{1}{2\kappa^2}\int_\Omega\sqrt{|g|}\left[R+\psi T-U\left(\psi\right)\right]d^5x+S_m\left(g_{ab},\phi\right).
\end{equation}
This time, it is the scalar field $\varphi$ that is absent from the action, both explicitly and via the potential $U$, which is now a function of solely $\psi$. One can thus obtain two equations of motion by taking the variation of $S$ with respect to $g_{ab}$ and $\psi$ respectively, which take the forms
\begin{eqnarray}\label{spec2_field}
R_{ab}\!-\!\frac{1}{2}g_{ab}\left(R\!+\!\psi T\!-\!U\right)\!=\!\kappa^2T_{ab}\!-\!\psi\left(T_{ab}\!+\!\Theta_{ab}\right),
\end{eqnarray}
\begin{equation}\label{spec2_kgphi}
U_\psi=T.
\end{equation}
Again, these two equations in particular could be obtained by imposing $U\left(\varphi,\psi\right)=U\left(\psi\right)$ and $\varphi=1$ in the general equations, see Eqs.\eqref{fieldst} and \eqref{eompsi}, but they are not a limit of the general case since Eq.~\eqref{eomphi} would force $R=0$, whereas here $R$ is not constrained. Taking a metric in the form of Eq.~\eqref{metricbrane} and the usual distribution of matter from Eq.~\eqref{actionchi}, the system of independent equations in this case takes the form
\begin{equation}\label{spec2_fieldsimp}
3A''=-\left(\frac{3}{2}\psi+2\right)\phi'^2-2\alpha\phi'\left(1+\psi\right),
\end{equation}
\begin{equation}\label{spec2_eqU}
U'=\big(\frac{3}{2}\phi'^2+5V-4\alpha\phi'\big)\psi',
\end{equation}
\begin{eqnarray}
&&\left(\frac{3}{4}\psi+1\right)\phi''+\left(\frac{3}{4}\psi'+\frac{3}{2}\psi A'+4A'\right)\phi'+\nonumber\\
&&+\alpha\left(4A'+4\psi A'+\psi'\right)=\left(\frac{5}{4}\psi+1\right)V_\phi.\label{spec2_eqA}
\end{eqnarray}
where Eq.~\eqref{spec2_fieldsimp} was obtained by subtracting the $(y,y)$ component from the $(t,t)$ component of Eq.~\eqref{spec2_field} and  Eq.~\eqref{spec2_eqU} was obtained vie the use of the chain rule on Eq.~\eqref{spec2_kgphi}. Similarly to the previous particular case, we note that even though we have one fewer equation and one fewer variable in comparison to the general case, there is also one fewer degree of freedom contained in the potential $U$ because it becomes a function of a single variable, and thus we can only impose two constraints to close the system instead of the three available in the general case. Again, we set the explicit form of $\phi$ and $V$, and leave $A$ to be computed from the equations. Our ansatz is the same as before, given by Eqs.\eqref{spec1_ansphi} to \eqref{spec1_ansW}, that is, again, we decided to keep the asymptotic baehavior of the model, instead of modifying the warp function. Inserting this ansatz into the system of Eqs.\eqref{spec2_fieldsimp} to \eqref{spec2_eqA}, one obtains a set of three coupled differential equations for $A$, $\psi$ and $U$ that must be integrated numerically subjected to a set of boundary conditions $A\left(y=0\right)=0$, $\psi\left(y=0\right)=\psi_0$, $U\left(y=0\right)=U_0$, and $\psi'\left(y=0\right)=0$.

\begin{figure}
\includegraphics[scale=0.8]{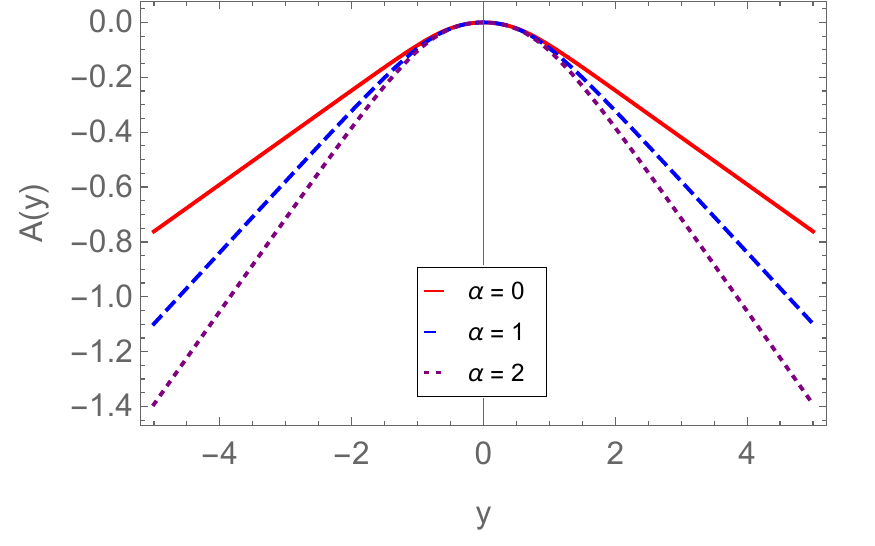}
\caption{Warp function $A\left(y\right)$ resulting from the system of Eqs.\eqref{spec2_fieldsimp} to \eqref{spec2_eqA} with $\psi_0=-1$, $\phi_0=U_0=k=1$ for different values of $\alpha$.}
\label{fig:spec2_warp}
\end{figure}

\begin{figure}
\includegraphics[scale=0.8]{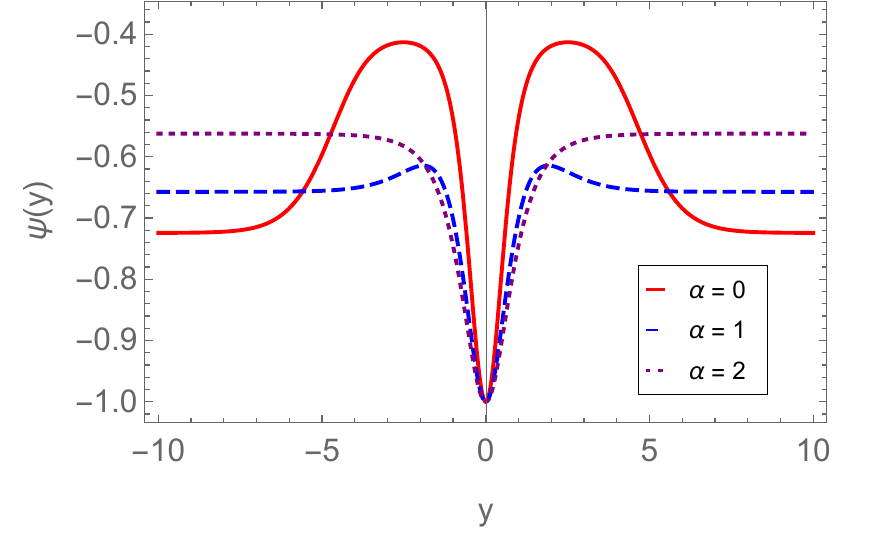}
\includegraphics[scale=0.8]{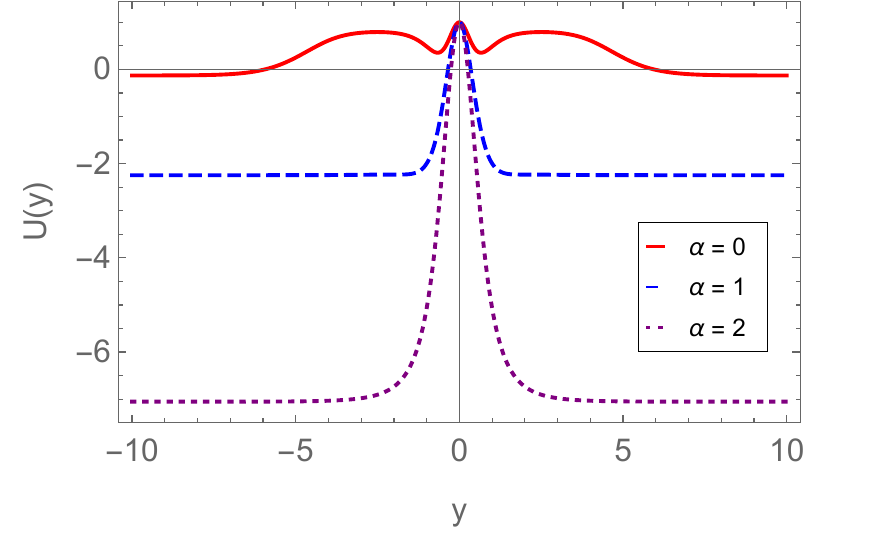}
\caption{Numerical solutions for $\varphi(y)$ (top panel) and $U(y)$ (bottom panel) from Eqs. \eqref{spec2_fieldsimp} to \eqref{spec2_eqA} with $\psi_0=-1$, $\phi_0=U_0=k=1$ for different values of $\alpha$.}
\label{fig:spec2_sols}
\end{figure}

In Fig.~\eqref{fig:spec2_warp} we plot the warp function $A\left(y\right)$ and in Fig.~\eqref{fig:spec2_sols} we plot the solutions for $\psi$ and $U$ for different values of $\alpha$. Again, one verifies that a variation in the Cuscuton parameter $\alpha$ affects the qualitative behavior of $\varphi$ and $U$, unlike it happens in the general case. In particular, an increase in the Cuscuton parameter changes the potential from a triple potential barrier to a single potential barrier at the brane, which affects the shape of $\varphi$ accordingly.

\begin{figure}
\includegraphics[scale=0.8]{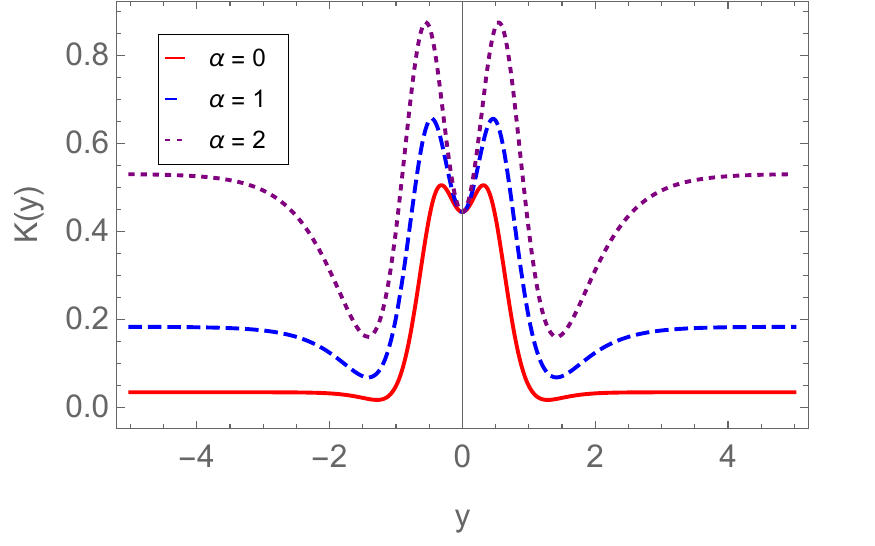}
    \caption{Numerical solutions for the Kretschmann scalar $K\left(y\right)$ from Eq.~\eqref{defKS} with $\psi_0=-1$, $\phi_0=U_0=k=1$ for different values of $\alpha$.}
    \label{fig:kretschmann2}
\end{figure}

Again, since the solution for the warp function $A$ is not standard but instead arises as a solution of the system of Eqs.\eqref{spec2_fieldsimp} to \eqref{spec2_eqA}, we have computed the Kretschmann scalar given in Eq.~\eqref{defKS} to guarantee that there are no unwanted divergences in the system. The results are plotter in Fig.~\ref{fig:kretschmann2}, and we see that it also behaves appropriately, for the values of the parameters used in this case.

\section{Stability}\label{secStability}

In this section we study the linear stability of the gravitational sector of the $f(R,T)-$brane in the scalar-tensor representation. For this purpose, we follow Ref. \cite{Bazeia:2015owa} and take linear perturbations in the scalar source field $\phi$ and the metric $g_{ab}$, so that $\phi\to\phi(y)+\delta\phi(r,y)$ and $g_{ab}\to g_{ab}(y)+\pi_{ab}(r,y)$, where $r$ represents the four-dimensional position vector and $\pi_{ab}$ is a symmetric tensor, with no component in extra dimension, and the four-dimensional components are represented by $\pi_{\mu\nu}=e^{2A}h_{\mu\nu}(r,y)$, where $h_{\mu\nu}(r,y)$ satisfies the transverse and traceless (TT) conditions $\partial^{\mu} h_{\mu\nu}=0$ and $h=\eta^{\mu\nu}h_{\mu\nu}=0$.

With this prescription, it is possible to show that if the potential $U$ in scalar-tensor representation is separable as $U_1(\varphi) +U_2(\psi)$, then the perturbation of the source scalar field decouples from the tensor perturbation of the metric. In this case we can obtain an equation for the tensor perturbation of metric in the form
\ben\label{stability1}
\Big(\!-\!\partial_y^2\! -\!4A^\prime \partial_y\!+\!e^{-2A}\Box^{(4)} \!-\!\frac{\varphi^\prime(y)}{\varphi(y)}\partial_y\!\Big)h_{\mu\nu}=0 \,.
\een

In order to better interpret this result, let us consider a  transformation of the form $dz=e^{-A(y)}dy$, which makes the metric conformally flat. In this case, the perturbed metric is written as
\begin{equation}\nonumber
ds^2=e^{2A(z)}\Big[(\eta_{\mu\nu}+h_{\mu\nu}(r,z))dx^\mu dx^\nu-dz^2\Big].
\end{equation}
Also, let us rewrite the tensor perturbation as $h_{\mu\nu}(r,z)=\xi(r)e^{-3A(z)/2}\varphi^{-1/2}{H}_{\mu\nu}$, where the function $\xi(r)$ obeys the plane wave equation $\Box^{(4)}\xi(r)=\omega^2\xi(r)$. Following these definitions, Eq.~ \eqref{stability1} can be written as a Schrödinger-like equation in the form
\ben\label{eq49}
\Big(-\frac{d^2}{dz^2}+{\cal U}(z)\Big){H}_{\mu\nu}= \omega^2{H}_{\mu\nu}\,,
\een
where the potential ${\cal U}(z)$ that governs the stability is written as,
\ben\label{eq50}
{\cal U}(z)=\alpha^2(z) -\frac{d\alpha}{dz}\,,
\een
where the function $\alpha(z)$ is defined as
\begin{equation}\nonumber
    \alpha(z)=-\frac32\frac{dA}{dz}-\frac12\frac{d}{dz}\big(\ln\varphi\,\big)\,.
\end{equation}

For this form, one verifies that the theory remains stable since Eq.~ \eqref{eq49} can be factorized in the form $S^{\cal y} S\,{H}_{\mu\nu}= \omega^2{H}_{\mu\nu}$, where $S^{\cal y}=-d/dz+\alpha(z)$, with ${\cal y}$ denoting hemitian conjugation, and $\omega^2\geq 0$. The massless graviton state represented by the zero mode is
\ben\label{eq51}
{H}^{(0)}(z)=N_0 \sqrt{\varphi(z)}\,e^{3A/2}\,,
\een
where $N_0$ is a normalization factor that can be computed via the integration of the zero mode
\ben\nonumber
N_0^2\int \varphi(y) e^{2A(y)}\,dy=1\,.
\een

\begin{figure}
\includegraphics[scale=0.8]{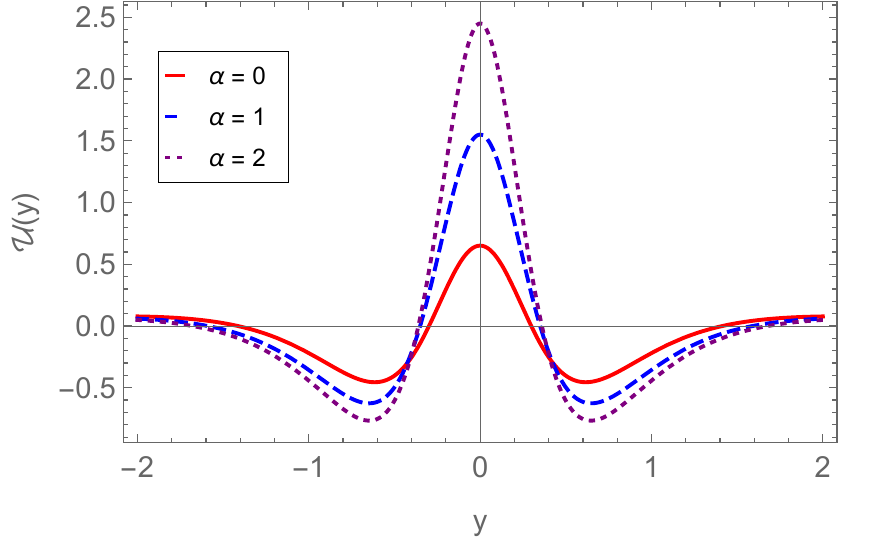}
\includegraphics[scale=0.8]{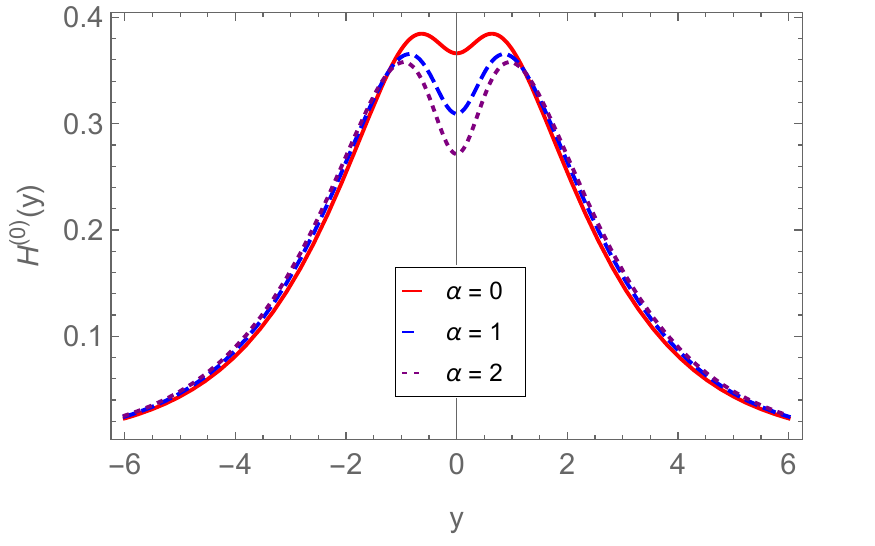}
\caption{Stability potential $\mathcal U(y)$ (top panel) and graviton zero-mode ${H}^{(0)}(y)$ (bottom panel) for the numerical solutions obtained in Sec.~\ref{sec:numerical1}, with $\varphi_0=10=-\psi_0$, $U_0=k=\phi_0=1$, $A_0=4/9$, and $A_1=-1/9$, for different values of $\alpha$.}
\label{fig:stabnum1}
\end{figure}

\begin{figure}
\includegraphics[scale=0.8]{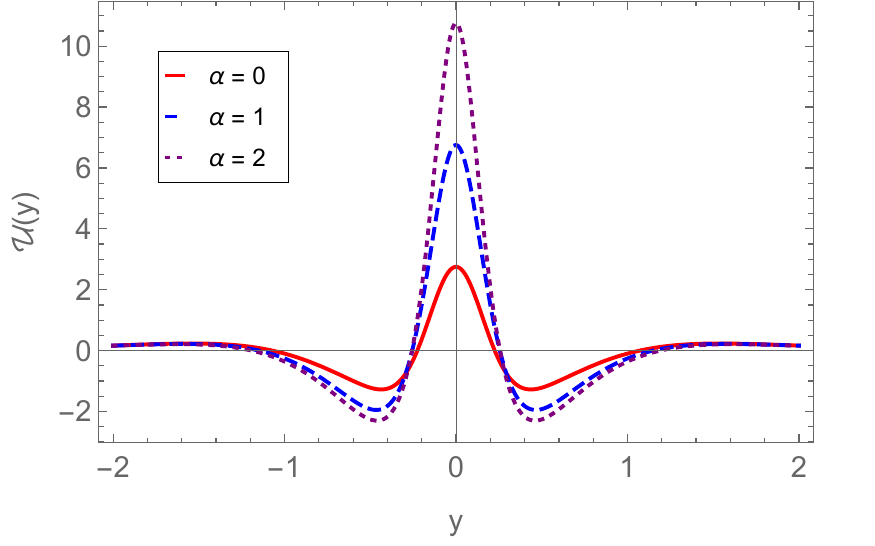}
\includegraphics[scale=0.8]{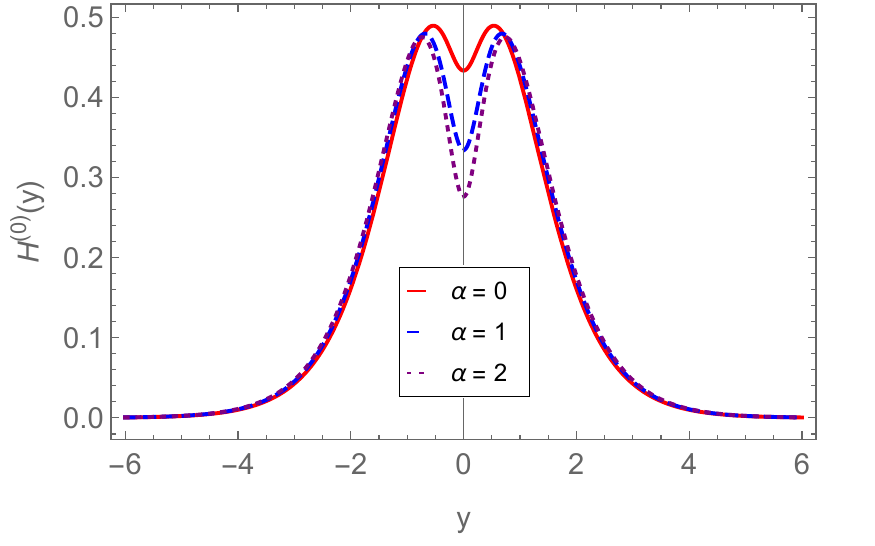}
\caption{Stability potential $\mathcal U(y)$ (top panel) and graviton zero-mode ${H}^{(0)}(y)$ (bottom panel) for the numerical solutions obtained in Sec.~\ref{sec:numerical2}, with $\varphi_0=U_0=k=\phi_0=A_0=1$, $\psi=-5$, for different values of $\alpha$.}
\label{fig:stabnum2}
\end{figure}

Let us use this formalism to analyze first the stability potential $\mathcal U$ and the graviton-zero mode ${H}^{(0)}$ of the solutions obtained numerically in Sec.~\ref{sec:numerical1}, which are plotted in Fig.~\ref{fig:stabnum1}. It is clear that even though the boundary conditions and free parameters were tuned in such a way as to guarantee that the graviton-zero mode of brane features an internal structure, the Cuscuton term contributes to an increase in the height of the potential barrier, resulting in an increase in the depth of the minimum of the graviton zero-mode and thus emphasizing the separation of the double-peak structure. The situation is qualitatively the same for the model presented in Sec.~\ref{sec:numerical2}, for which the stability potential $\mathcal U$ and the graviton zero-mode ${H}^{(0)}$ are plotted in Fig.~ \ref{fig:stabnum2}.

\begin{figure}
\includegraphics[scale=0.8]{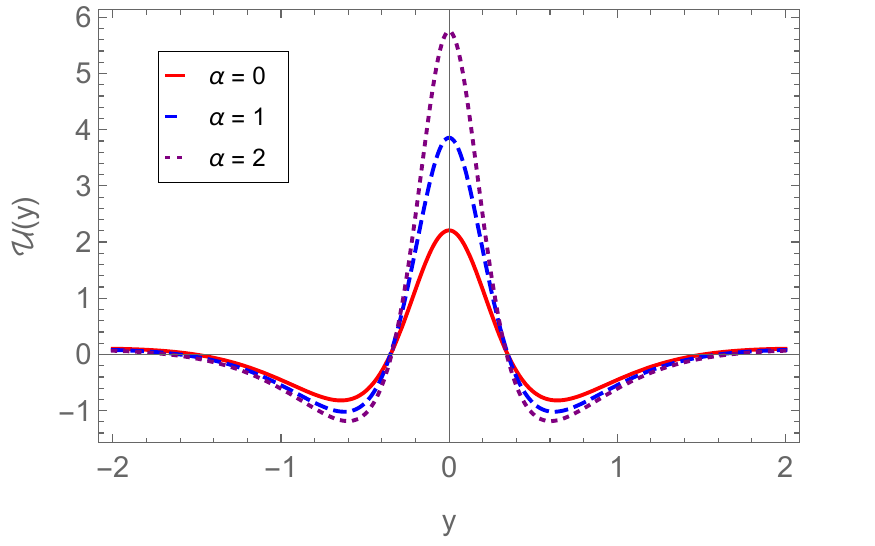}
\includegraphics[scale=0.8]{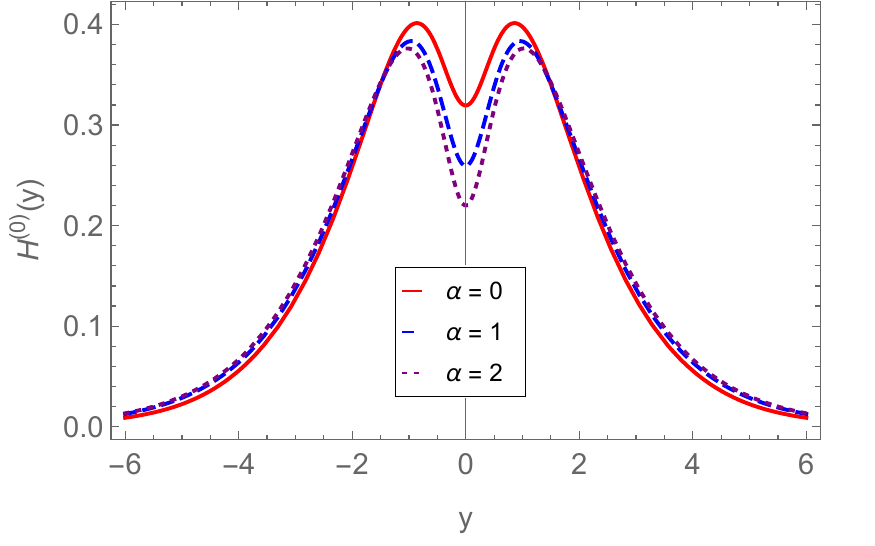}
\caption{Stability potential $\mathcal U(y)$ (top panel) and graviton zero-mode ${H}^{(0)}(y)$ (bottom panel) for the numerical solutions obtained in Sec.~\ref{sec:spec1}, with $\varphi_0=-1$, $U_0=k=\phi_0=1$, for different values of $\alpha$.}
\label{fig:stabspec1}
\end{figure}

\begin{figure}
\includegraphics[scale=0.8]{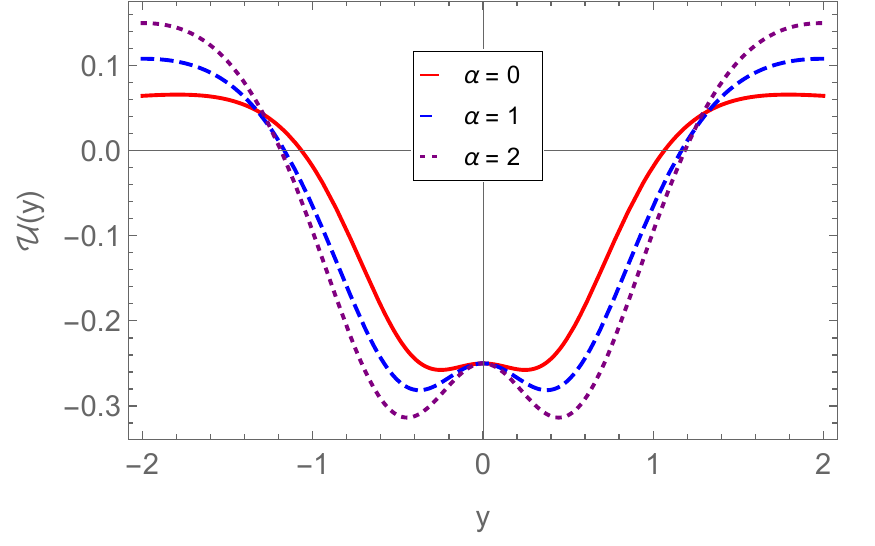}
\includegraphics[scale=0.8]{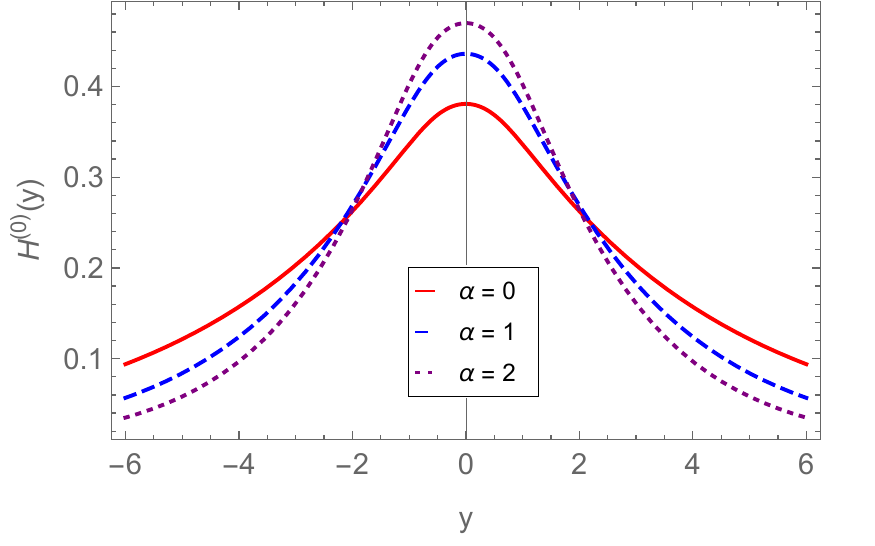}
\caption{Stability potential $\mathcal U(y)$ (top panel) and graviton zero-mode ${H}^{(0)}(y)$ (bottom panel) for the numerical solutions obtained in Sec.~\ref{sec:spec2}, with $\psi_0=-1$, $U_0=k=1$, $A_0=0$ for different values of $\alpha$.}
\label{fig:stabspec2}
\end{figure}

In what comes to the particular cases studied in Sec.~\ref{sec:spec1} and \ref{sec:spec2}, there are a few comments to outline. For the particular case $f\left(R,T\right)=F\left(R\right)+T$, the stability potential $\mathcal U$ and the graviton zero-mode ${H}^{(0)}$ are plotted in Fig.~ \ref{fig:stabspec1}. In these figures, it is clearly visible that even though a single scalar degree of freedom of the theory is present, i.e., the scalar field $\varphi$, the addition of the Cuscuton term allows one to find solutions with internal structure in zero-mode, a situation that was unattainable when the source field was assumed to have standard dynamics, see Ref. \cite{Rosa:2021tei}. However, the same is not true for the particular case $f\left(R,T\right)=R+G(T)$. In this situation, no matter the choice of free parameters and the value of $\alpha$ the stability potential is always a double potential well, and thus the graviton zero-mode never develops a double-peak structure, i.e., the brane never develops internal structure, as we can see in Fig.~\ref{fig:stabspec2}.

\section{Comments and conclusions}\label{sec:conclusion}

In this work, we have extender our previous studies of the $f(R,T)-$brane models in the scalar-tensor representation of the theory. We have obtained mostly numerical solutions for the auxiliary fields, using kink-like solutions as ansatz for the source field of the brane. The main improvement in comparison to our previous works is the addition of the Cuscuton term in action source of the brane, for which we analyzed the influence on general case and particular separable cases.

We started by using numerical methods to investigate the differential equations arising from the more general brane model in the scalar-tensor representation, with the source field of brane having dynamics modified by inclusion of the Cuscuton term. In this situation we have shown that the inclusion of the Cuscuton term, although altering the asymptotic behavior of auxiliary fields and non-negligibly affecting the shape of the solutions, it does not introduce any other qualitative modification for the model other than an increase of the zero-mode split behavior of the gravitational sector stability. However, if one considers instead the particular cases for which a single auxiliary scalar field is present, i.e., the separable $f\left(R,T\right)=F\left(R\right)+T$ and $f\left(R,T\right)=R+G\left(T\right)$, the effects of adding a Cuscuton term become much more evident, inclusively altering the general behavior of the scalar field potential $U$ and, consequently, the scalar fields themselves (either $\varphi$ or $\psi$ respectively, depending on the particular situation considered).

A particularly interesting result concerns the addition of the Cuscuton term to the situation where only the scalar field $\varphi$ is present. The Cuscuton term allows one to induce an internal structure on the brane even in this situation, whereas in a previous work \cite{Rosa:2021tei} we have found that if the source field presents standard dynamics then no internal structure can be induced with the scalar field $\varphi$ only. This means that the Cuscuton dynamics also plays a role in this case, contributing to change the geometry of the brane, an issue that may modify phenomenological aspects of particle physics in the $F(R)+T$ scenario. This motivates other investigations, in particular, on the entrapment of fermions \cite{Cruz:2011ru,Li:2017dkw,Mazani:2020abe}. Moreover, it may also contribute to change the cosmological behavior under possible braneworld cosmology scenarios; see, e.g., Refs. \cite{Langlois:2000iu,Bazeia:2007vx} and references therein.

The same effect is not present for the model with the scalar field $\psi$ alone, for which we have shown that, no matter the intensity of the Cuscuton term, the graviton zero-mode is always single-peaked on the brane. Thus, it seems like the scalar field associated to the arbitrary dependence of the function $f\left(R,T\right)$ in $T$ is \textit{per se} not fundamental in the development of an internal structure, whereas the scalar field associated with the dependence in $R$ is essential. This information is of current interest, and may guide us toward studies concerned with the importance of the Cuscuton within the scalar-tensor representation of $f(R,T)$ gravity.

We verified that despite the modifications introduced in this work, the robustness of the brane model has remained unchanged. This indicates that the modifications are of current interest and can be used to analyze other  situations as in the recent study \cite{Bazeia:2021jok} where the authors add the Cuscuton term in source two-field models that led to new and interesting split behaviors in the warp factor. Following this, an immediate question to address in future works would be to analyze whether the split warp factor behavior is maintained in generalized two-field brane models with scalar-tensor representation. Another issue concerning two-field scenarios is the possibility recently considered in \cite{Gauy:2022xsc}, where one also adds another extra dimension, to search for intersecting thick brane configurations in a braneworld environment with two extra dimensions, in $(5,1)$ dimensional spacetime. This is of interest since the enlarged scenario may perhaps give rise to more accurate phenomenology, which motivates the introduction of the Cuscuton dynamics. We can also focus on the possibility of finding domain wall solutions from cosmological solutions, searching for first order equations induced on the brane \cite{Bazeia:2007vx}, within the scalar-tensor representation of $f(R,T)$ gravity and, in particular, within the $F(R)+T$ decomposition and the addition of the Cuscuton term. Distinct possibilities explored in \cite{XY1,XY2} are also of current interest; for instance, in \cite{XY1} the authors considered the case of $f(R,T^2)$, with $T^2=T_{\mu\nu}T^{\mu\nu}$, and in \cite{XY2} the modification used $f(R,T,R_{\mu\nu}T^{\mu\nu})$. We think that such distinct scenarios could also be investigated in the presence of a warped geometry with a single extra dimension of infinite extent in the presence of the Cuscuton term. These and other issues are under investigation and we hope to report on some on them in a future work.

\begin{acknowledgments}

DB would like to thank CNPq (Brazil), grants No. 404913/2018-0 and No. 303469/2019-6, and Paraiba State research foundation, FAPESQ-PB, grant No. 0015/2019, for partial financial support. JLR was supported by the European Regional Development Fund and the programme Mobilitas Pluss (MOBJD647).

\end{acknowledgments}



\begin{thebibliography}{99}

\bibitem{fr}T. P. Sotiriou and V. Faraoni, Rev. Mod. Phys. 82, 451 (2010).
\bibitem{Capozziello:2011et}
S.~Capozziello and M.~De Laurentis,
Phys. Rept. \textbf{509}, 167-321 (2011)
doi:10.1016/j.physrep.2011.09.003
[arXiv:1108.6266 [gr-qc]].

\bibitem{Nojiri:2010wj}
S.~Nojiri and S.~D.~Odintsov,
Phys. Rept. \textbf{505}, 59-144 (2011)
doi:10.1016/j.physrep.2011.04.001
[arXiv:1011.0544 [gr-qc]].


\bibitem{Solodukhin:1994pc}
S.~N.~Solodukhin,
Phys. Rev. D \textbf{51} (1995), 591-602
doi:10.1103/PhysRevD.51.591
[arXiv:hep-th/9405132 [hep-th]].

\bibitem{Rippl:1995bg}
S.~Rippl, H.~van Elst, R.~K.~Tavakol and D.~Taylor,
Gen. Rel. Grav. \textbf{28} (1996), 193-205
doi:10.1007/BF02105423
[arXiv:gr-qc/9511010 [gr-qc]].

\bibitem{Noh:1996da}
H.~Noh and J.~C.~Hwang,
Phys. Rev. D \textbf{55} (1997), 5222-5225
doi:10.1103/PhysRevD.55.5222
[arXiv:gr-qc/9610059 [gr-qc]].


\bibitem{Charmousis:2002rc}
C.~Charmousis and J.~F.~Dufaux,
Class. Quant. Grav. \textbf{19} (2002), 4671-4682
doi:10.1088/0264-9381/19/18/304
[arXiv:hep-th/0202107 [hep-th]].

\bibitem{Hwang:2001pu}
J.~C.~Hwang and H.~Noh,
Phys. Lett. B \textbf{506} (2001), 13-19
doi:10.1016/S0370-2693(01)00404-X
[arXiv:astro-ph/0102423 [astro-ph]].

\bibitem{DeFelice:2010aj}
A.~De Felice and S.~Tsujikawa,
Living Rev. Rel. \textbf{13} (2010), 3
doi:10.12942/lrr-2010-3
[arXiv:1002.4928 [gr-qc]].


\bibitem{Harko:2011kv}
T.~Harko, F.~S.~N.~Lobo, S.~Nojiri and S.~D.~Odintsov,
Phys. Rev. D \textbf{84} (2011), 024020
doi:10.1103/PhysRevD.84.024020
[arXiv:1104.2669 [gr-qc]].

\bibitem{Cai:2015emx}
Y.~F.~Cai, S.~Capozziello, M.~De Laurentis and E.~N.~Saridakis,
Rept. Prog. Phys. \textbf{79} (2016) no.10, 106901
doi:10.1088/0034-4885/79/10/106901
[arXiv:1511.07586 [gr-qc]].

\bibitem{K1}C. Armendariz-Picon, T. Damour, V.F. Mukhanov, Phys. Lett. B458 (1999) 209, hep-th/9904075.
\bibitem{K2}J. Garriga , V.F. Mukhanov, Phys. Lett. B458 (1999) 219, hep-th/9904176.
\bibitem{K3}T. Chiba, T. Okabe, M. Yamaguchi, Phys. Rev. D62 (2000) 023511, astro-ph/9912463.
\bibitem{K4}C. Armendariz-Picon, V.F. Mukhanov, P.J. Steinhardt, Phys. Rev. Lett. 85 (2000) 4438, astro-ph/0004134.
\bibitem{K5}E. Babichev, Phys. Rev. D74 (2006) 085004, hep-th/0608071 
\bibitem{Adam:2007ag}
C.~Adam, N.~Grandi, J.~Sanchez-Guillen and A.~Wereszczynski,
J. Phys. A \textbf{41} (2008), 212004
[erratum: J. Phys. A \textbf{42} (2009), 159801]
doi:10.1088/1751-8113/41/21/212004
[arXiv:0711.3550 [hep-th]].

\bibitem{Barton}B. Zwiebach, {\it A First Course in String Theory.} 2nd Edition, Cambridge University Press, 2009
\bibitem{Randall:1999vf}
L.~Randall and R.~Sundrum,
Phys. Rev. Lett. \textbf{83}, 4690-4693 (1999)
doi:10.1103/PhysRevLett.83.4690
[arXiv:hep-th/9906064 [hep-th]].

\bibitem{Goldberger:1999uk}
W.~D.~Goldberger and M.~B.~Wise,
Phys. Rev. Lett. \textbf{83}, 4922-4925 (1999)
doi:10.1103/PhysRevLett.83.4922
[arXiv:hep-ph/9907447 [hep-ph]].

\bibitem{Skenderis:1999mm}
K.~Skenderis and P.~K.~Townsend,
Phys. Lett. B \textbf{468}, 46-51 (1999)
doi:10.1016/S0370-2693(99)01212-5
[arXiv:hep-th/9909070 [hep-th]].

\bibitem{DeWolfe:1999cp}
O.~DeWolfe, D.~Z.~Freedman, S.~S.~Gubser and A.~Karch,
Phys. Rev. D \textbf{62}, 046008 (2000)
doi:10.1103/PhysRevD.62.046008
[arXiv:hep-th/9909134 [hep-th]].

\bibitem{Csaki:2000fc}
C.~Csaki, J.~Erlich, T.~J.~Hollowood and Y.~Shirman,
Nucl. Phys. B \textbf{581}, 309-338 (2000)
doi:10.1016/S0550-3213(00)00271-6
[arXiv:hep-th/0001033 [hep-th]].

\bibitem{Afonso:2007gc}
V.~I.~Afonso, D.~Bazeia, R.~Menezes and A.~Y.~Petrov,
Phys. Lett. B \textbf{658}, 71-76 (2007)
doi:10.1016/j.physletb.2007.10.038
[arXiv:0710.3790 [hep-th]].

\bibitem{Bazeia:2013oha}
D.~Bazeia, R.~Menezes, A.~Y.~Petrov and A.~J.~da Silva,
Phys. Lett. B \textbf{726} (2013), 523-526
doi:10.1016/j.physletb.2013.08.068
[arXiv:1306.1847 [hep-th]].

\bibitem{Bazeia:2013uva}
D.~Bazeia, A.~S.~Lob\~ao, Jr., R.~Menezes, A.~Y.~Petrov and A.~J.~da Silva,
Phys. Lett. B \textbf{729} (2014), 127-135
doi:10.1016/j.physletb.2014.01.011
[arXiv:1311.6294 [hep-th]].


\bibitem{Bazeia:2015dna}
D.~Bazeia, A.~Lobao, L.~Losano, R.~Menezes and A.~Y.~Petrov,
Phys. Rev. D \textbf{92} (2015) no.6, 064010
doi:10.1103/PhysRevD.92.064010
[arXiv:1502.02564 [hep-th]].

\bibitem{Correa:2015qma}
R.~A.~C.~Correa and P.~H.~R.~S.~Moraes,
Eur. Phys. J. C \textbf{76}, no.2, 100 (2016)
doi:10.1140/epjc/s10052-016-3952-9
[arXiv:1509.00732 [hep-th]].

\bibitem{Correa:2015ako}
R.~A.~C.~Correa, P.~H.~R.~S.~Moraes, A.~de Souza Dutra and R.~da Rocha,
Phys. Rev. D \textbf{92}, no.12, 126005 (2015)
doi:10.1103/PhysRevD.92.126005
[arXiv:1508.04493 [hep-th]].

\bibitem{Rosa:2020uli}
J.~L.~Rosa, D.~A.~Ferreira, D.~Bazeia and F.~S.~N.~Lobo,
Eur. Phys. J. C \textbf{81} (2021) no.1, 20
doi:10.1140/epjc/s10052-021-08840-3
[arXiv:2010.10074 [gr-qc]].


\bibitem{Bazeia:2015owa}
D.~Bazeia, A.~S.~Lob\~ao and R.~Menezes,
Phys. Lett. B \textbf{743} (2015), 98-103
doi:10.1016/j.physletb.2015.02.037
[arXiv:1502.04757 [hep-th]].

\bibitem{Zhong}Y. Zhong,Y. Zhong,Y. P. Zhang,Y. X. Liu, 
Eur. Phys. J. C 78, no. 1, 45 (2018)
arXiv:1711.09413 [hep-th]

\bibitem{Guo:2018tpo}
W.~D.~Guo, Y.~Zhong, K.~Yang, T.~T.~Sui and Y.~X.~Liu,
Phys. Lett. B \textbf{800}, 135099 (2020)
doi:10.1016/j.physletb.2019.135099
[arXiv:1805.05650 [hep-th]].

\bibitem{Sui:2021uic}
T.~T.~Sui, Y.~P.~Zhang, B.~M.~Gu and Y.~X.~Liu,
Eur. Phys. J. C \textbf{81}, no.11, 980 (2021)
doi:10.1140/epjc/s10052-021-09756-8

\bibitem{Bazeia:2021jok}
D.~Bazeia, D.~A.~Ferreira and M.~A.~Marques,
Eur. Phys. J. C \textbf{81} (2021) no.7, 619
doi:10.1140/epjc/s10052-021-09434-9
[arXiv:2102.06932 [hep-th]].

\bibitem{Rosa:2021tei}
J.~L.~Rosa, M.~A.~Marques, D.~Bazeia and F.~S.~N.~Lobo,
[arXiv:2105.06101 [gr-qc]]. Eur. Phys J. C 81, 981 (2021).

\bibitem{Rosa:2021teg}
J.~L.~Rosa,
Phys. Rev. D \textbf{103} (2021) no.10, 104069
doi:10.1103/PhysRevD.103.104069
[arXiv:2103.11698 [gr-qc]].

\bibitem{CG1}N. Afshordi, D. J. Chung and G. Geshnizjani,
Cuscuton: A Causal Field Theory with an
Infinite Speed of Sound, Phys. Rev. D 75 (2007) 083513 [hep-th/0609150].
\bibitem{CG2}N. Afshordi, D. J. Chung, M. Doran and G. Geshnizjani, Cuscuton Cosmology: Dark Energy
meets Modified Gravity, Phys. Rev. D 75 (2007) 123509 [astro-ph/0702002].

\bibitem{CC01}S. S. Boruah, H. J. Kim and G. Geshnizjani,
Theory of Cosmological Perturbations with
Cuscuton, JCAP 07 (2017) 022 [1704.01131].



\bibitem{CC02}S. S. Boruah, H. J. Kim, M. Rouben and G. Geshnizjani, Cuscuton bounce, JCAP 08 (2018) 031 [1802.06818].

\bibitem{Ito:2019fie}
A.~Ito, A.~Iyonaga, S.~Kim and J.~Soda,
Phys. Rev. D \textbf{99} (2019) no.8, 083502
doi:10.1103/PhysRevD.99.083502
[arXiv:1902.08663 [astro-ph.CO]].

\bibitem{Andrade:2018afh}
I.~Andrade, M.~A.~Marques and R.~Menezes,
Nucl. Phys. B \textbf{942}, 188-204 (2019)
doi:10.1016/j.nuclphysb.2019.03.016
[arXiv:1806.01923 [hep-th]].

\bibitem{Afshordi:2006ad}
N.~Afshordi, D.~J.~H.~Chung and G.~Geshnizjani,
Phys. Rev. D \textbf{75} (2007), 083513
doi:10.1103/PhysRevD.75.083513
[arXiv:hep-th/0609150 [hep-th]].

\bibitem{Afshordi:2007yx}
N.~Afshordi, D.~J.~H.~Chung, M.~Doran and G.~Geshnizjani,
Phys. Rev. D \textbf{75} (2007), 123509
doi:10.1103/PhysRevD.75.123509
[arXiv:astro-ph/0702002 [astro-ph]].

\bibitem{Afshordi:2009tt}
N.~Afshordi,
Phys. Rev. D \textbf{80} (2009), 081502
doi:10.1103/PhysRevD.80.081502
[arXiv:0907.5201 [hep-th]].

\bibitem{Bazeia:2016}
D. Bazeia, Elisama E.M. Lima, L. Losano
Eur. Phys. J. C 77 (2017) 127
doi: 10.1140/epjc/s10052-017-4701-4
[arXiv:1611.09314 [hep-th]].

\bibitem{Cruz:2011ru}
W.~T.~Cruz, A.~R.~Gomes and C.~A.~S.~Almeida,
Eur. Phys. J. C \textbf{71}, 1790 (2011)
doi:10.1140/epjc/s10052-011-1790-3
[arXiv:1110.4651 [hep-th]].

\bibitem{Li:2017dkw}
Y.~Y.~Li, Y.~P.~Zhang, W.~D.~Guo and Y.~X.~Liu,
Phys. Rev. D \textbf{95}, no.11, 115003 (2017)
doi:10.1103/PhysRevD.95.115003
[arXiv:1701.02429 [hep-th]].

\bibitem{Mazani:2020abe}
E.~Mazani, A.~Tofighi and M.~M.~Sorkhi,
Eur. Phys. J. C \textbf{80}, no.3, 267 (2020)
doi:10.1140/epjc/s10052-020-7826-9


\bibitem{Langlois:2000iu}
D.~Langlois, R.~Maartens, M.~Sasaki and D.~Wands,
Phys. Rev. D \textbf{63}, 084009 (2001)
doi:10.1103/PhysRevD.63.084009
[arXiv:hep-th/0012044 [hep-th]].

\bibitem{Bazeia:2007vx}
D.~Bazeia, F.~A.~Brito and F.~G.~Costa,
Phys. Lett. B \textbf{661}, 179-185 (2008)
doi:10.1016/j.physletb.2008.02.016
[arXiv:0707.0680 [hep-th]].

\bibitem{Gauy:2022xsc}
H.~M.~Gauy and A.~E.~Bernardini,
Phys. Rev. D \textbf{105}, no.2, 024068 (2022)
doi:10.1103/PhysRevD.105.024068
[arXiv:2201.01284 [hep-th]].

\bibitem{XY1}N. Katırcı and M. Kavuk, Eur. Phys. J. Plus 129, 163 (2014) doi:10.1140/epjp/i2014-14163-6 [arXiv:1302.4300 [gr-qc]].
\bibitem{XY2}Z. Haghani et. al., Phys. Rev. D 88, no.4, 044023 (2013)
doi:10.1103/PhysRevD.88.044023 [arXiv:1304.5957 [gr-qc]].


\end{thebibliography}
\end{document}